\documentclass[11pt]{article}
\usepackage[a4paper,margin=25mm]{geometry}
\usepackage{amsmath,amssymb,bm}
\usepackage{graphicx,booktabs,multirow,array}
\usepackage{xcolor}
\usepackage{listings}
\usepackage{tikz}
\usetikzlibrary{arrows.meta,positioning,fit,calc}
\usepackage{algorithm,algpseudocode}
\usepackage[numbers,sort&compress]{natbib}
\usepackage[colorlinks=true,linkcolor=blue!50!black,citecolor=blue!50!black,urlcolor=blue!50!black]{hyperref}

\newcommand{\R}{\mathbf{R}}
\newcommand{\bu}{\mathbf{u}}
\newcommand{\bpsi}{\bm{\psi}}

\title{CuAD: Source-to-Source Automatic Differentiation\\of CUDA Kernels for Discrete Adjoints\\in Computational Fluid Dynamics}
\author{Seunghyun Joo\\[2pt]
\small Institute of Advanced Aerospace Technology, Seoul National University\\
\small Seoul 08826, Republic of Korea\\
\small \texttt{chlrh45351@gmail.com}}
\date{}

\begin{document}
\maketitle

\begin{abstract}
Discrete adjoint methods give the sensitivities that aerodynamic design and data assimilation
require, but they remain hard to obtain for flow solvers that run on graphics processors. This
paper presents CuAD, a source-to-source automatic differentiation tool that reads CUDA kernels and
writes their tangent and adjoint versions as ordinary CUDA source. Its reverse mode is organised
around the GPU thread. It adopts the convention of CUDA kernels that inputs and outputs do not alias,
recomputes intermediate values instead of storing them, differentiates pure accumulations without a tape
and keeps any remaining tape in a fixed thread-local array. We evaluate CuAD on four kernels from simple
flow models and on a two-dimensional Euler solver. All generated derivatives pass dot-product tests to
round-off, and the adjoint kernels cost two to six times the primal kernels. On a Biot--Savart summation
compiled by one compiler, the CuAD adjoint costs about four times the primal at every size, as does Enzyme
when the source declares the absence of aliasing with \texttt{\_\_restrict\_\_}. On the unchanged source
Enzyme caches the loop in the device heap and costs 9 to 240 times the primal, while the heap tape of Clad
costs 60 to 4400 times the primal and fails beyond 8192 vortices. Exact Jacobians assembled from the
generated tangents also matter for accuracy. In single precision, the adjoint gradient of a
finite-difference Jacobian reaches errors of 45 to 75 per cent on a stiff Burgers problem and of
$5\times10^{-4}$ to $4\times10^{-2}$ on a transonic airfoil, whereas the generated Jacobian keeps them
below $5\times10^{-5}$ and $3\times10^{-6}$. The applications include the shape sensitivities of lift and drag
of a transonic NACA~0012, for which the residual, the boundary conditions and the wall force are
differentiated by CuAD and the adjoint agrees with the tangent to $10^{-13}$.
\end{abstract}


% =====================================================================
\section{Introduction}
\label{sec:intro}

Adjoint methods made gradient-based design practical in fluid dynamics because their cost does not
grow with the number of design variables~\cite{jameson1988,giles2000}. The discrete variant linearises
the residual of the solver itself, so its gradient is consistent with the discrete solution that the
optimiser sees~\cite{peter2010}. This consistency requires a derivative for every term of the
residual, and mature solvers obtain these derivatives with automatic differentiation
(AD)~\cite{griewank2008}. Source transformation with Tapenade~\cite{hascoet2013} underlies the ADjoint
approach~\cite{mader2008}, whereas operator overloading with CoDiPack~\cite{sagebaum2019} underlies the
discrete adjoint of SU2~\cite{economon2016,albring2016}. Kenway et al.\ review these choices for computational fluid
dynamics~\cite{kenway2019}.

Graphics processors have moved many flow solvers onto thousands of concurrent threads. A steady
solution on a modest grid now takes seconds, so an optimisation loop can afford many flow and adjoint
solutions, provided that the adjoint itself runs on the same hardware. The derivative tools above,
however, were designed for sequential CPU code, and their central mechanism does not carry over. A
reverse sweep must restore the intermediate values of the forward computation, and on a CPU these values
are kept on a tape. On a GPU the memory available to each thread is small, and thousands of threads that
record and replay a tape concurrently must synchronise, which makes taping
costly~\cite{gremse2016}. The reverse sweep also turns
concurrent reads of shared data into concurrent writes of adjoints, which must either be proved free of
conflicts or be protected with atomic operations~\cite{huckelheim2022toms,huckelheim2022icpp}.

The tools that now differentiate GPU code address these problems at a general level.
Section~\ref{sec:related} compares them in detail. Frameworks such as Warp and JAX-Fluids differentiate
programs written in their own language, so an existing CUDA solver would have to be rewritten. Enzyme
acts on the compiler intermediate representation, where the derivative is no longer visible as source
code~\cite{moses2020,moses2021}. Clad, a Clang plugin, has recently added reverse mode for CUDA kernels
and kernel launches~\cite{vassilev2015,ifrim2022,koutsou2025}. It keeps a tape inside every thread. Its authors
report that redundant tape operations severely degraded the performance of the generated gradients until
the tape was reworked, and they list an analysis that reduces tape storage to the minimum as future
work~\cite{koutsou2025}. Enzyme decides between storing and recomputing a value from what its alias
analysis can prove, so the cost of its adjoint depends on information that a CUDA kernel rarely states
explicitly. Neither tool exploits what is known about the kernels of a flow solver in particular.

This structure is the starting point of the present work. Almost every kernel of a finite-difference or
finite-volume solver belongs to one of two families. In a stencil kernel each thread gathers a few
neighbouring values and writes the residual of one cell. In a summation kernel each thread accumulates a
long sum over other cells, panels or particles, as in vortex methods, panel methods and integral
boundary conditions. Both families read their inputs and write their outputs in separate arrays, by a
convention so common that it is rarely written down. Under this convention every intermediate value of a
thread can be recomputed from the inputs in the reverse sweep, and the old value of an accumulator is
never needed at all. A reverse mode designed around the GPU thread can therefore avoid storing anything
for these kernels, and its cost no longer grows with the length of the loops inside a thread.

A second issue concerns the accuracy of GPU adjoints. Many GPU solvers compute in single precision and
form the Jacobians they need for implicit time stepping and adjoint solutions by finite differences of
their residual kernel. When the Jacobian is approximate, the error of the adjoint solution is governed by
the conditioning of the transposed Jacobian, and Dwight and Brezillon showed that simplified Jacobians can
mislead an optimiser~\cite{dwight2006}. The error that single precision adds to a finite-difference
Jacobian cannot be removed by any choice of step, but how far it propagates into adjoint gradients has not
been quantified. A Jacobian assembled from generated tangents by graph colouring~\cite{curtis1974} has no
truncation error, so the comparison isolates the effect of the finite differences.

This paper presents CuAD, a source-to-source AD tool that writes the tangent and adjoint of CUDA kernels as
readable CUDA, and uses it to build complete discrete adjoints of flow problems on the GPU. The
contributions are the following.
\begin{enumerate}
\item \textbf{A thread-centred reverse mode for flow-solver kernels.} Under the convention that a kernel's
inputs and outputs do not alias, CuAD recomputes intermediate values in the reverse sweep, differentiates pure
accumulations without a tape and keeps the scalars that a loop carries from one iteration to the next in a
fixed thread-local array. Adjoint updates are atomic only for memory that several threads can reach. For
the kernels studied here no tape is needed at all.
\item \textbf{A controlled comparison with Clad and Enzyme.} On a Biot--Savart summation compiled by one
compiler the CuAD adjoint costs about four times the primal at every size up to 16384 vortices. Enzyme
reaches the same cost once the absence of aliasing is declared with \texttt{\_\_restrict\_\_}, which shows
that the convention, not the tool, decides the cost. On the unchanged source Enzyme caches the loop in the
device heap and Clad keeps a heap tape, and their costs grow to two and three orders of magnitude above the
primal with up to 8~GB of heap.
\item \textbf{A quantification of the error of single-precision adjoints.} With all linear systems solved in
double precision, the adjoint gradient of a single-precision finite-difference Jacobian reaches errors of
45 to 75 per cent on a stiff Burgers problem and of 0.05 to 4 per cent on a transonic airfoil, while the
Jacobian assembled from generated tangents in the same precision keeps it three orders of magnitude lower.
\item \textbf{Complete GPU discrete adjoints built only from generated kernels.} For a transonic NACA~0012 the
residual, the wall and far-field boundary conditions and the wall force are differentiated by CuAD without
change to the source, and the shape sensitivities of lift and drag from the adjoint agree with those from the
tangent to $10^{-13}$. Further applications cover a viscous shock, a scalar sensor in a vortex, the inverse
design of a nozzle and the unsteady adjoint of a rolling-up vortex sheet.
\end{enumerate}

Section~\ref{sec:related} reviews related tools. Section~\ref{sec:adjoint} summarises the discrete
adjoint and the ways in which a solver uses generated derivatives. Section~\ref{sec:design} describes
CuAD. Section~\ref{sec:problems} defines the four model problems and their kernels, which
Section~\ref{sec:verif} verifies. Section~\ref{sec:clad} compares CuAD with Clad and Enzyme and
Section~\ref{sec:exact} studies inexact Jacobians. Section~\ref{sec:apps} presents the applications,
Section~\ref{sec:limits} states the limitations and Section~\ref{sec:concl} concludes.

% =====================================================================
\section{Related work}
\label{sec:related}

AD tools fall into three groups according to the level at which they act on a program. Source
transformation reads the program text and writes a new program that computes the derivatives. Tapenade
is the most widely used tool of this kind for Fortran and C~\cite{hascoet2013}, and its generated code
is the basis of several adjoint flow solvers~\cite{mader2008}. Because the derivative is ordinary source,
it can be compiled with the same optimisations as the primal code and inspected by the developer.
Operator overloading replaces the arithmetic types of a program by types that record every operation on
a tape at run time and replay it backwards~\cite{griewank2008}. CoDiPack is a modern C++ example with
expression templates~\cite{sagebaum2019}. The approach needs no parser, but the tape must be stored and
traversed, which on a GPU is limited by the memory available per thread and by the synchronisation of
concurrent tape writes~\cite{gremse2016}. Compiler-based tools
act on an intermediate representation inside a compiler. Enzyme differentiates LLVM IR after
optimisation, which allows it to apply GPU-specific transformations such as caching decisions and the
removal of redundant atomic operations~\cite{moses2020,moses2021}.

Parallel code adds the problem of write conflicts in the reverse sweep. Hückelheim and Hascoët extended
Tapenade to OpenMP worksharing loops~\cite{huckelheim2022toms} and later used formal methods to prove,
from the memory access pattern of the primal loop, that some adjoint writes cannot conflict, which lets
them drop atomic updates or private copies~\cite{huckelheim2022icpp}. Clad is a source transformation
tool that runs as a Clang plugin~\cite{vassilev2015}. It first differentiated CUDA device functions
called from kernels~\cite{ifrim2022} and now differentiates kernels, kernel launches and the memory management of
the host code that calls them~\cite{koutsou2025}. Clad marks every adjoint write to a pointer parameter
as atomic and reports that redundant tape operations initially dominated the overhead of its gradients
on the LULESH hydrodynamics benchmark~\cite{koutsou2025}.

Outside general-purpose AD, several frameworks offer derivatives for particular programming models. NVIDIA
Warp generates forward and adjoint code for simulation kernels written in a subset of
Python~\cite{warp2022}. JAX-Fluids writes a compressible two-phase flow solver entirely in JAX and
obtains gradients through the automatic differentiation of JAX~\cite{bezgin2023}. For energies defined on meshes, a locality-aware
system assembles gradients and sparse Hessians with per-element forward mode in registers and shared
memory, without a global tape~\cite{mahmoud2026}. These frameworks require the solver to be written in
their language or model. CuAD instead targets existing CUDA C++ kernels. Table~\ref{tab:tools} summarises
the features that matter for discrete adjoints of GPU flow solvers.

\begin{table}[ht]
\centering
\caption{Features of AD tools relevant to discrete adjoints of GPU flow solvers.}
\label{tab:tools}
\resizebox{\textwidth}{!}{%
\begin{tabular}{lllllll}
\toprule
Tool & Level & Input & CUDA reverse mode & Derivative output & Concurrent adjoint writes & Reduced storage \\
\midrule
Tapenade~\cite{hascoet2013,huckelheim2022icpp} & source & Fortran, C & partial & source & proof-based (parallel loops) & TBR analysis \\
CoDiPack~\cite{sagebaum2019} & overloading & C++ & -- & tape at run time & -- & -- \\
Enzyme~\cite{moses2021} & LLVM IR & any LLVM language & device functions & compiled code & compiler analysis & cache optimisation \\
Clad~\cite{koutsou2025} & Clang plugin & C++, CUDA & kernels and launches & compiled or dumped source & atomic for pointer parameters & listed as future work \\
Warp~\cite{warp2022} & JIT & Python subset & kernels & compiled code & framework arrays & last-write rule \\
JAX-Fluids~\cite{bezgin2023} & JAX AD & JAX & -- & XLA program & -- & checkpointing \\
CuAD (this work) & source & CUDA C++ & kernels and device functions & readable CUDA source & atomic for shared memory only & accumulations untaped \\
\bottomrule
\end{tabular}}
\end{table}

% =====================================================================
\section{Discrete adjoints with generated kernels}
\label{sec:adjoint}

\subsection{Steady problems}

Let the discrete residual $\R(\bu,\bm\alpha)\in\mathbb R^m$ define the state $\bu$ for design variables
$\bm\alpha$ through $\R(\bu,\bm\alpha)=\mathbf 0$, and let $J(\bu,\bm\alpha)$ be an objective. The
total derivative of $J$ follows from the adjoint equation
\begin{equation}
\left(\frac{\partial\R}{\partial\bu}\right)^{\!\mathsf T}\bpsi = -\left(\frac{\partial J}{\partial\bu}\right)^{\!\mathsf T},
\qquad
\frac{\mathrm dJ}{\mathrm d\bm\alpha} = \frac{\partial J}{\partial\bm\alpha} + \bpsi^{\mathsf T}\frac{\partial\R}{\partial\bm\alpha}.
\label{eq:adj}
\end{equation}
A GPU solver evaluates $\R$ with kernels in which each thread computes the residual of one cell. AD of
such a kernel provides two products. The tangent kernel returns the Jacobian--vector product
$(\partial\R/\partial\bu)\dot\bu+(\partial\R/\partial\bm\alpha)\dot{\bm\alpha}$, and the adjoint kernel
returns the transposed products $(\partial\R/\partial\bu)^{\mathsf T}\bar{\R}$ and
$(\partial\R/\partial\bm\alpha)^{\mathsf T}\bar{\R}$ for a seed $\bar{\R}$.

Both products enter the adjoint method. The design term of Eq.~\eqref{eq:adj} is a single call of the
adjoint kernel with the seed $\bar\R=\bpsi$. The adjoint equation itself can be solved without the
Jacobian by a Krylov method that calls the adjoint kernel once per iteration, or it can be solved with
the assembled Jacobian.

\subsection{Jacobian assembly by colouring}

A stencil residual has a sparse Jacobian whose columns can be grouped so that no two columns of a group
share a non-zero row~\cite{curtis1974,gebremedhin2005}. If $\mathbf s_c$ denotes the sum of the unit
vectors of the columns in group $c$, the tangent product
\begin{equation}
\mathbf t_c = \frac{\partial\R}{\partial\bu}\,\mathbf s_c
\label{eq:colour}
\end{equation}
contains each non-zero of those columns exactly once, so the entry $(i,j)$ of column $j$ in group $c$ is
the $i$-th component of $\mathbf t_c$. A three-point stencil needs three groups, a five-point stencil on
a two-dimensional grid needs five groups with the colour $(i+2j)\bmod 5$, and a system with $n_v$
variables per cell multiplies the count by $n_v$. The number of tangent evaluations therefore depends on
the stencil only, not on the grid size, and the assembled Jacobian is exact up to the rounding of the
tangent arithmetic.

\subsection{Unsteady problems}

For a time integration $\bu^{n+1}=\bm\Phi(\bu^n,\bm\alpha)$ with an objective $J(\bu^{N})$ at the final
step, the discrete adjoint runs backwards in time,
\begin{equation}
\bar\bu^{N} = \left(\frac{\partial J}{\partial\bu^N}\right)^{\!\mathsf T},\qquad
\bar\bu^{n} = \left(\frac{\partial\bm\Phi}{\partial\bu^n}\right)^{\!\mathsf T}\bar\bu^{n+1},\qquad
\frac{\mathrm dJ}{\mathrm d\bm\alpha} = \sum_{n=0}^{N-1}\left(\frac{\partial\bm\Phi}{\partial\bm\alpha}\right)^{\!\mathsf T}\bar\bu^{n+1}.
\label{eq:unsteady}
\end{equation}
Each transposed product is a call of the adjoint kernels that make up one step of $\bm\Phi$, evaluated at
the stored state $\bu^n$. Section~\ref{sec:vortex} applies this recursion to a two-stage scheme.

% =====================================================================
\section{Design of CuAD}
\label{sec:design}

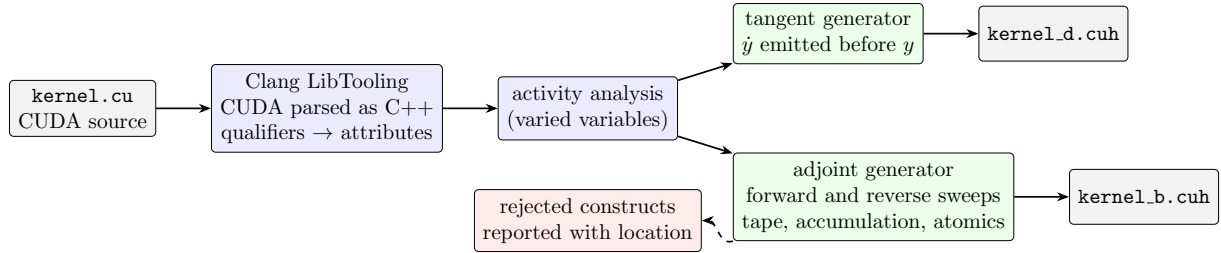
\begin{figure}[ht]
\centering
\resizebox{\textwidth}{!}{%
\begin{tikzpicture}[node distance=6mm and 9mm, every node/.style={font=\small},
  box/.style={draw, rounded corners=2pt, align=center, minimum height=9mm, inner sep=4pt, fill=#1},
  box/.default=white, arr/.style={-{Stealth[length=2mm]}, thick}]
\node[box=gray!10] (in) {\texttt{kernel.cu}\\CUDA source};
\node[box=blue!8, right=of in] (parse) {Clang LibTooling\\CUDA parsed as C++\\qualifiers $\to$ attributes};
\node[box=blue!8, right=of parse] (act) {activity analysis\\(varied variables)};
\node[box=green!8, above right=2mm and 9mm of act] (fwd) {tangent generator\\$\dot y$ emitted before $y$};
\node[box=green!8, below right=2mm and 9mm of act] (rev) {adjoint generator\\forward and reverse sweeps\\tape, accumulation, atomics};
\node[box=gray!10, right=of fwd] (outd) {\texttt{kernel\_d.cuh}};
\node[box=gray!10, right=of rev] (outb) {\texttt{kernel\_b.cuh}};
\node[box=red!8, below=8mm of act] (diag) {rejected constructs\\reported with location};
\draw[arr] (in) -- (parse); \draw[arr] (parse) -- (act);
\draw[arr] (act) -- (fwd); \draw[arr] (act) -- (rev);
\draw[arr] (fwd) -- (outd); \draw[arr] (rev) -- (outb);
\draw[arr, dashed] (rev.south west) to[out=200,in=0] (diag.east);
\end{tikzpicture}}
\caption{Structure of CuAD. The two generators write ordinary CUDA source; constructs they cannot
differentiate are collected as located diagnostics.}
\label{fig:pipeline}
\end{figure}

Figure~\ref{fig:pipeline} shows the structure of the tool. A CUDA file is parsed into a Clang abstract
syntax tree, an activity analysis marks the variables that carry derivatives, and one of two generators
writes the tangent or the adjoint companion of every function with a floating-point interface. The
constructs that the generators cannot differentiate are reported instead of being translated.

\subsection{Parsing CUDA without a GPU toolchain}

CuAD is built on Clang LibTooling~\cite{clang}. Rather than invoking the CUDA front end, it parses a
\texttt{.cu} file as C++ in which \texttt{\_\_global\_\_} and \texttt{\_\_device\_\_} are mapped to
\texttt{annotate} attributes. The built-in index variables and the math library are declared in a small
in-memory header, and the remaining CUDA-only tokens are defined away. The resulting syntax tree contains
every arithmetic expression of a kernel together with the information needed to restore its qualifiers.
Because no GPU back end is involved, the tool runs on any machine with a Clang development installation.
Expressions of the primal code are copied verbatim from the source buffer, so the generated code stays
close to what the developer wrote and can be read line by line against it.

\subsection{Activity analysis}

A variable is varied when it depends, directly or through a chain of assignments, on an active
floating-point input. CuAD computes this property with a fixed-point iteration over the assignments of
each function. A call to a user function marks as varied only those arguments that the callee can write,
namely pointers, arrays and references, so a scalar passed by value to a library function stays passive.
Only varied variables receive derivative storage. The analysis follows dependencies forward from the
inputs.

\subsection{Tangent mode}

For a function \texttt{f}, the tangent companion \texttt{f\_d} takes a direction \texttt{x\_d} next to
every active argument \texttt{x}. Each statement is preceded by its tangent statement, so the tangent
reads the primal values that hold before the statement overwrites them. A call to a user function is
replaced by a call to its tangent companion, which computes the primal value as well. When such a call
is nested inside a larger expression, as in \texttt{0.5*(c(a)+c(b))}, the companion is called before the
statement and the expression uses the returned tangent. The tangent kernel of the Euler residual in
Appendix~\ref{app:listings} shows both forms.

\subsection{Reverse mode for GPU threads}

The adjoint companion \texttt{f\_b} runs the primal statements forward and then the adjoint statements in
reverse order, accumulating input adjoints as
$\bar{\mathbf x}\mathrel{+}=(\partial\mathbf f/\partial\mathbf x)^{\mathsf T}\bar{\mathbf y}$. Four rules
adapt this pattern to a GPU thread.

\paragraph{Inputs and outputs do not alias.} The reverse sweep reads the inputs of a kernel again, after the
forward sweep has written its outputs. CuAD therefore assumes that the arrays a kernel reads are not
overwritten by the kernel itself, which is the convention of residual and summation kernels and the
property that a \texttt{\_\_restrict\_\_} qualifier declares. Under this convention every value that a thread
computes from its inputs can be recomputed instead of stored. A kernel that updates an input array in place
violates the convention and must be split into a kernel that reads and a kernel that writes.

\paragraph{Recompute rather than store.} A loop whose body declares local variables is recomputed iteration
by iteration in the backward loop, so its locals are never stored. A scalar declared outside such a loop and
reassigned inside it is carried from one iteration to the next. The forward sweep saves its value at the
start of every iteration and the backward loop restores it before recomputing, so these checkpoints are the
only values a loop stores. They are kept on a per-thread stack that is a fixed local array of the generated
function, sized at compile time, rather than in the device heap. A function that saves nothing does not
declare the stack.

\paragraph{Accumulations need no tape.} A carried scalar that is only accumulated, $v\leftarrow v+e$, needs no
checkpoint. The derivative of the new $v$ with respect to the old $v$ is one, so the reverse sweep never needs
the old value as long as every other read of $v$ follows all of its accumulations and does not lie in a loop
that also accumulates into it. CuAD recognises this pattern and saves nothing for it. Without the analysis a
summation over $N$ elements would require a stack of $2N$ entries per thread to be reserved at compile time,
even though the values stored in it are never read.

\paragraph{Atomic updates only where threads can collide.} An adjoint that lives in a pointer parameter
refers to memory that many threads may update at once, so CuAD accumulates into it with
\texttt{atomicAdd}. Adjoints of thread-local data use a plain addition. A value that the primal loads into
a local variable collects all its contributions locally and reaches shared memory in a single atomic
update, while an array element read directly in several expressions receives one update per occurrence.
Parameters declared as arrays, such as \texttt{double f[3]}, are treated as thread-local.

\paragraph{Control flow.} Counted loops are reversed, including grid-stride loops. Conditions of \texttt{if} statements are
evaluated again in the reverse sweep, which requires that a branch does not reassign a variable its own
condition reads. Algorithm~\ref{alg:reverse} summarises the generation of a reverse sweep.

\begin{algorithm}[ht]
\caption{Generation of the adjoint companion of a function body $B$ (simplified).}
\label{alg:reverse}
\begin{algorithmic}[1]
\State $V \gets$ varied variables of $B$ \Comment{activity analysis}
\State $A \gets$ variables of $V$ assigned only as $v\leftarrow v+e$ and read elsewhere only after these updates
\State $T \gets \{v\in V\setminus A : v$ is a reassigned local scalar$\}$ \Comment{taped variables}
\State emit the forward sweep of $B$, with \texttt{push}$(v)$ before each assignment to $v\in T$
\State \quad and, in a loop that declares locals, \texttt{push} of the carried $v\in T$ at the top of each iteration
\For{each statement $s$ of $B$ in reverse order}
  \If{$s$ is a counted loop} emit the reversed loop; if it declares locals, \texttt{pop} the carried $v\in T$ and recompute its body
  \ElsIf{$s$ is a conditional} emit the condition and the reversed branches
  \ElsIf{$s$ assigns $e$ to $v\in A$} propagate $\bar v$ to the operands of $e$; keep $\bar v$
  \ElsIf{$s$ assigns $e$ to $v$} \texttt{pop}$(v)$ if $v\in T$; propagate $\bar v$ to the operands of $e$; $\bar v\gets 0$
  \Else\ record an \emph{unsupported construct} at the location of $s$
  \EndIf
\EndFor
\State write each operand adjoint to its location: \texttt{atomicAdd} for shared memory, $+$ for thread-local data
\end{algorithmic}
\end{algorithm}

Listing~\ref{lst:bs} shows the summation kernel used in Section~\ref{sec:clad} and the core of its
adjoint. The accumulators \texttt{su} and \texttt{sv} belong to the set $A$ of Algorithm~\ref{alg:reverse},
so the reverse loop recomputes \texttt{dx}, \texttt{dy}, \texttt{r2} and \texttt{k} and stores nothing.

\begin{lstlisting}[language=C++,caption={Biot--Savart summation (top) and the reverse loop that CuAD generates (bottom, abridged).},label={lst:bs}]
for (int j = 0; j < n; j++) {
  double dx = xi - xs[j];  double dy = yi - ys[j];
  double r2 = dx*dx + dy*dy + delta2;  double k = g[j] / (6.283185307179586 * r2);
  su = su - k*dy;  sv = sv + k*dx;
}
// ---- generated reverse loop ----
for (int j = (n) - 1; j >= ((int)(0)); j--) {
  double dx = xi - xs[j];  double dy = yi - ys[j];      // recomputed, not restored
  double r2 = dx*dx + dy*dy + delta2;  double k = g[j] / (6.283185307179586 * r2);
  double dx_b = 0, dy_b = 0, r2_b = 0, k_b = 0;
  k_b += (sv_b)*(dx);  dx_b += (k)*(sv_b);  k_b += (-(su_b))*(dy);  dy_b += (k)*(-(su_b));
  atomicAdd(&g_b[j], (k_b)/((6.283185307179586 * r2)));
  r2_b += (6.283185307179586)*(-((k_b)*(g[j]))/(((6.283185307179586*r2))*((6.283185307179586*r2))));
  dx_b += (r2_b)*(dx);  dx_b += (dx)*(r2_b);  dy_b += (r2_b)*(dy);  dy_b += (dy)*(r2_b);
  yi_b += dy_b;  atomicAdd(&ys_b[j], -(dy_b));  xi_b += dx_b;  atomicAdd(&xs_b[j], -(dx_b));
}
\end{lstlisting}

\subsection{Rejecting what cannot be differentiated}

A wrong derivative that looks plausible is more harmful to an optimisation than no derivative, because the
optimiser follows it without warning. CuAD therefore refuses to generate code for any construct whose
derivative it cannot guarantee, such as loops without a fixed trip count that touch active data or calls to
functions without a derivative rule, and reports the location of the construct instead.

% =====================================================================
\section{Model problems and kernels}
\label{sec:problems}

Four model problems provide the kernels of the evaluation. Each represents an operation that is common in
flow solvers, and each kernel follows the usual GPU layout in which one thread computes the residual of one
cell or the velocity at one point. The problems are simple enough for the derivatives to be checked
independently, yet they contain the constructs that make AD of real solvers difficult: nonlinear fluxes,
device functions with array outputs, value-returning functions inside expressions and long per-thread
summations.

\subsection{K1: steady viscous Burgers equation}

The viscous Burgers equation
\begin{equation}
u\,\frac{\partial u}{\partial x} = \nu\,\frac{\partial^2 u}{\partial x^2} + f(x),\qquad x\in[0,1],
\end{equation}
is the simplest model of a viscous shock. On a uniform grid $x_i=ih$, $h=1/(n-1)$, the kernel evaluates
the central-difference residual
\begin{equation}
R_i = u_i\,\frac{u_{i+1}-u_{i-1}}{2h} - \nu\,\frac{u_{i+1}-2u_i+u_{i-1}}{h^2} - f_i,\qquad 0<i<n-1,
\label{eq:k1}
\end{equation}
whose Jacobian is tridiagonal with entries
$\partial R_i/\partial u_{i\mp1}=\mp u_i/2h-\nu/h^2$ and $\partial R_i/\partial u_i=(u_{i+1}-u_{i-1})/2h+2\nu/h^2$.
The benchmark uses $4.2\times10^{6}$ cells. In Sections~\ref{sec:exact} and~\ref{sec:burgersapp} the residual
is multiplied by $h$, which is its finite-volume form, and closed by $u_0=1$ and $u_{n-1}=-0.6$.

\subsection{K2: two-dimensional advection--diffusion}

A passive scalar $q$ transported by a divergence-free velocity $\mathbf v=(a,b)$ with diffusivity $\kappa$
obeys
\begin{equation}
\nabla\cdot(\mathbf v q - \kappa\nabla q) = s .
\end{equation}
On a uniform Cartesian grid of cells $(i,j)$ the kernel evaluates the finite-volume flux balance
\begin{equation}
R_{ij} = h\left(F_{i+\frac12,j}-F_{i-\frac12,j}+G_{i,j+\frac12}-G_{i,j-\frac12}\right),\qquad
F_{i+\frac12,j} = \bar a_{i+\frac12,j}\,\frac{q_{ij}+q_{i+1,j}}{2} - \kappa\,\frac{q_{i+1,j}-q_{ij}}{h},
\label{eq:k2}
\end{equation}
with face velocities $\bar a_{i+\frac12,j}=(a_{ij}+a_{i+1,j})/2$ and analogous expressions for $G$ in the
$y$-direction. The residual is linear in $q$ and bilinear in $(q,\mathbf v)$, so the kernel tests the
differentiation with respect to the velocity field as well. The benchmark uses $2048^2$ cells and a
divergence-free cellular velocity field.

\subsection{K3: quasi-one-dimensional Euler equations}

The flow through a nozzle of slowly varying cross-section $A(x)$ is described by
\begin{equation}
\frac{\partial\mathbf W}{\partial t} + \frac{\partial\mathbf F}{\partial x} = \mathbf S,\qquad
\mathbf W = \begin{pmatrix}\rho A\\ \rho uA\\ EA\end{pmatrix},\quad
\mathbf F = \begin{pmatrix}\rho uA\\ (\rho u^2+p)A\\ (E+p)uA\end{pmatrix},\quad
\mathbf S = \begin{pmatrix}0\\ p\,\mathrm dA/\mathrm dx\\ 0\end{pmatrix},
\end{equation}
with $p=(\gamma-1)(E-\tfrac12\rho u^2)$ and $\gamma=1.4$. The kernel evaluates the finite-volume residual with
central fluxes and scalar artificial dissipation scaled by the spectral radius $\lambda=|u|+c$,
\begin{equation}
\mathbf R_i = \frac{\mathbf H_{i+\frac12}-\mathbf H_{i-\frac12}}{h} - \mathbf S_i,\qquad
\mathbf H_{i+\frac12} = \tfrac12(\mathbf F_i+\mathbf F_{i+1}) - \varepsilon\,\bar\lambda_{i+\frac12}(\mathbf W_{i+1}-\mathbf W_i),
\label{eq:k3}
\end{equation}
where $\bar\lambda_{i+\frac12}=\tfrac12(\lambda_i+\lambda_{i+1})$, $\varepsilon=0.25$ and
$\mathrm dA/\mathrm dx$ is a central difference. The flux $\mathbf F$ and the spectral radius $\lambda$ are
device functions; the flux writes its three components into a local array and the spectral radius returns
a value used inside a sum. The benchmark uses $1.05\times10^{6}$ cells.

\subsection{K4: Biot--Savart summation over point vortices}

Vortex methods represent a flow by $N$ point vortices with circulations $\Gamma_j$ at positions
$\mathbf x_j=(x_j,y_j)$. With the regularisation of Krasny~\cite{krasny1986}, the velocity induced at
$\mathbf x_i$ is
\begin{equation}
u_i = -\sum_{j=1}^{N}\frac{\Gamma_j}{2\pi}\,\frac{y_i-y_j}{|\mathbf x_i-\mathbf x_j|^2+\delta^2},\qquad
v_i = \sum_{j=1}^{N}\frac{\Gamma_j}{2\pi}\,\frac{x_i-x_j}{|\mathbf x_i-\mathbf x_j|^2+\delta^2}.
\label{eq:k4}
\end{equation}
Each thread evaluates the velocity at one vortex by a loop over all $N$ vortices, so the per-thread work
and the length of any tape grow with $N$. The benchmark uses $N=8192$; the tool comparison varies $N$ from
256 to 16384.

% =====================================================================
\section{Verification}
\label{sec:verif}

For each kernel the tangent is compared with a central finite difference of the primal kernel in double
precision along a random direction $\dot{\mathbf x}$. The adjoint is checked with the dot-product identity
$\langle\bar{\mathbf y},J\dot{\mathbf x}\rangle=\langle J^{\mathsf T}\bar{\mathbf y},\dot{\mathbf x}\rangle$ for a
random seed $\bar{\mathbf y}$, which tests the adjoint against the tangent to round-off independently of any
step size. Timings are medians of 50 launches.

Table~\ref{tab:verif} shows that all tangents agree with the finite differences to the accuracy of the
difference itself and that all adjoints satisfy the dot-product identity to round-off. The tangent costs
between 2.2 and 2.4 times the primal kernel. The adjoint costs between 2.1 and 6.3 times the primal; the
largest ratio belongs to the Euler residual, whose adjoint recomputes the device functions of the flux and
the spectral radius for three neighbouring cells. None of the adjoints uses a tape or local memory.

\begin{table}[ht]
\centering
\caption{Verification and cost of the generated kernels in double precision.}
\label{tab:verif}
\resizebox{\textwidth}{!}{%
\begin{tabular}{lcccccc}
\toprule
Kernel & Tangent vs.\ FD & Dot test & Primal (ms) & Tangent/primal & Adjoint/primal & Adjoint registers \\
\midrule
K1 Burgers 1-D        & $5.5\times10^{-11}$ & $4.3\times10^{-16}$ & 0.27 & 2.28 & 3.03 & 37 \\
K2 Adv.--diff.\ 2-D   & $2.1\times10^{-11}$ & $2.6\times10^{-16}$ & 0.70 & 2.25 & 2.07 & 54 \\
K3 Euler quasi-1-D    & $1.2\times10^{-9}$  & $2.4\times10^{-16}$ & 0.54 & 2.42 & 6.26 & 86 \\
K4 Biot--Savart       & $2.4\times10^{-9}$  & $1.9\times10^{-16}$ & 3.78 & 2.38 & 4.05 & 50 \\
\bottomrule
\end{tabular}}
\end{table}

These ratios fix the cost of the adjoint computations in Section~\ref{sec:apps}. A Krylov solution of the
adjoint equation that calls the adjoint kernel once per iteration costs between two and six residual
evaluations per iteration, which is comparable to the cost of one iteration of an implicit flow solver.
Because none of the adjoints stores a tape, their memory footprint equals that of the primal kernel plus
one adjoint array for each active array, independent of the length of the loops inside a thread.

% =====================================================================
\section{Comparison with Clad and Enzyme}
\label{sec:clad}

\subsection{Set-up}

Clad~2.0 and Enzyme~0.0.186 were built against the same Clang and LLVM~19, and all adjoints were compiled by
that Clang with the same options and run on the same GPU, so that the comparison concerns the generated code
rather than the compilers. Clad emits a gradient as a device function, and Enzyme differentiates a device
function through a call to \texttt{\_\_enzyme\_autodiff}; each was launched from a minimal kernel, and the CuAD
adjoint kernel was launched directly. Every adjoint was compared with the CuAD adjoint for the same seed, which
in turn passes the dot-product test of Section~\ref{sec:verif}. Clad and Enzyme store their tape or cache in
the device heap, whose size must be fixed before the first launch. Their timings use a heap of 8~GB, and the
smallest heap with which they run was determined separately by doubling from 64~MB. Each configuration was
timed in three separate processes, and the median of these runs is reported with their range.

\subsection{Stencil kernels}

Table~\ref{tab:clad} shows that the two tools agree to $10^{-16}$ on the Burgers and advection--diffusion
kernels and that their adjoints take the same time. Neither stores a tape for these kernels, their register
counts differ by at most four, and the generated code has the same structure. For stencil residuals the two
approaches are equivalent.

\begin{table}[ht]
\centering
\caption{CuAD and Clad adjoints of the stencil kernels compiled by the same compiler. Times are medians of 50
launches. The Biot--Savart kernel is compared in Table~\ref{tab:bs}.}
\label{tab:clad}
\resizebox{\textwidth}{!}{%
\begin{tabular}{lcccccc}
\toprule
& & \multicolumn{2}{c}{Adjoint / primal} & \multicolumn{2}{c}{Registers / stack (B)} & \\
Kernel & Agreement & CuAD & Clad & CuAD & Clad & Clad tape \\
\midrule
K1 Burgers 1-D      & $1.4\times10^{-17}$ & 3.16 & 3.15 & 38 / 0 & 38 / 0 & none \\
K2 Adv.--diff.\ 2-D & $7.0\times10^{-17}$ & 1.99 & 1.99 & 50 / 0 & 46 / 0 & none \\
K3 Euler quasi-1-D  & --                  & --   & \multicolumn{3}{c}{Clad terminates while differentiating} & -- \\
\bottomrule
\end{tabular}}
\end{table}

\subsection{Summation kernel}

\begin{figure}[ht]
\centering
\includegraphics[width=\textwidth]{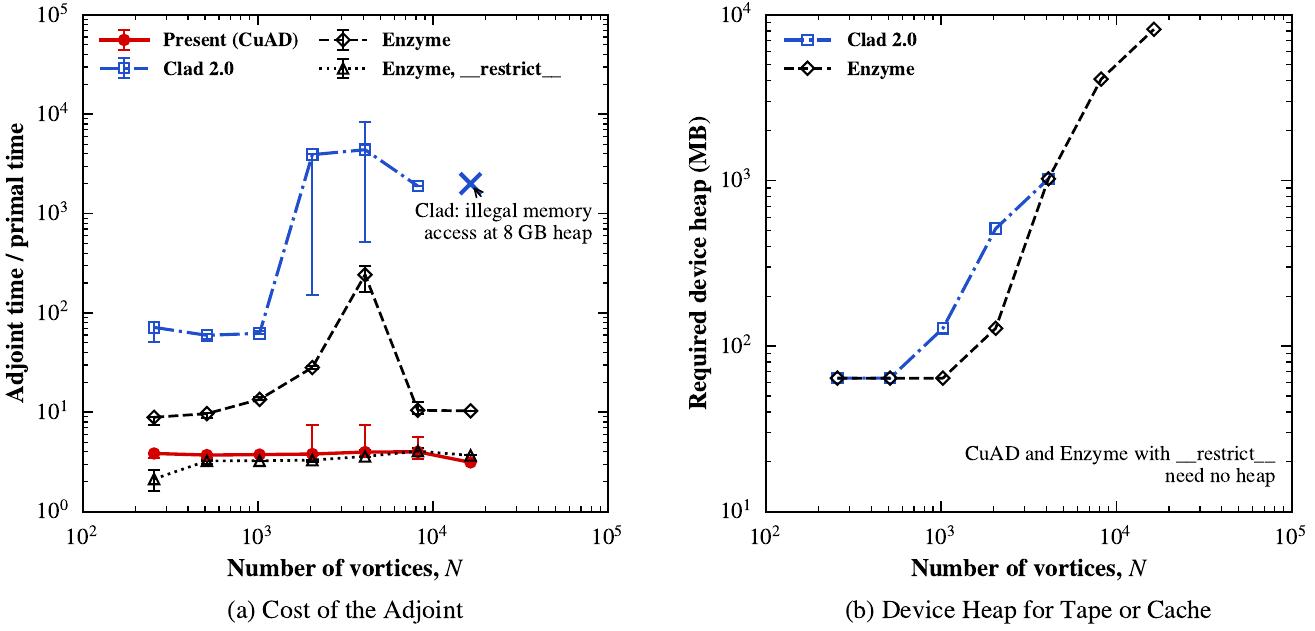}
\caption{Biot--Savart adjoint of four tools compiled by one compiler. (a) Median cost of three runs relative to
the primal kernel, with the range of the runs, at an 8~GB heap; the cross marks the size at which every Clad
run failed. (b) Smallest device heap, found by doubling from 64~MB, with which the heap-based adjoints run.}
\label{fig:bs}
\end{figure}

The Biot--Savart kernel shows where the approaches differ. Figure~\ref{fig:bs} and Table~\ref{tab:bs} report
four adjoints of the same source. The CuAD adjoint recomputes \texttt{dx}, \texttt{dy}, \texttt{r2} and
\texttt{k} in the backward loop and stores nothing (Listing~\ref{lst:bs}). Its cost stays between 3.1 and
4.0 times the primal for every $N$ from 256 to 16384, it uses no local memory and no heap, and it agrees with
the others to $4\times10^{-15}$.

Enzyme reaches the same cost when the pointer arguments of the differentiated function are declared
\texttt{\_\_restrict\_\_}. With this declaration its alias analysis proves that the loop reads arrays nothing
writes, so it recomputes the loop as CuAD does, at 2.1 to 4.1 times the primal and without a heap. Without the
declaration, which is how the kernel is written, Enzyme cannot exclude that the outputs \texttt{uind} and
\texttt{vind} overwrite the inputs. It therefore caches the values of every iteration in memory allocated from
the device heap. The required heap grows from 64~MB to 8~GB with $N$, and the adjoint costs 9 to 28 times the
primal up to $N=2048$, 240 times at $N=4096$ and 10 times beyond. The peak at $N=4096$ recurred in all three
runs and in the search for the smallest heap.

Clad stores \texttt{dx}, \texttt{dy}, \texttt{r2} and \texttt{k} on a tape in every iteration. Its GPU tape
keeps a small buffer inside the thread (2680~bytes of local memory) and grows into the device heap, so each
thread holds a tape whose length is proportional to $N$. The adjoint costs 60 to 70 times the primal up to
$N=1024$. Beyond that size its cost varies between runs by more than an order of magnitude, from 150 to 8300
times the primal, and one run in three stopped with an illegal memory access at $N=4096$ and at $N=8192$.
At $N=16384$ every run failed.

The comparison separates two effects. The convention that inputs and outputs do not alias, which CuAD adopts
and which \texttt{\_\_restrict\_\_} declares to Enzyme, is what allows the loop to be recomputed. Where the
values are stored instead, their location decides the cost: a tape in the device heap is shared by all
threads through a global allocator, while CuAD's stack is a fixed array in each thread's local memory.

\begin{table}[ht]
\centering
\caption{Biot--Savart adjoint at selected sizes: median cost relative to the primal over three runs (range in
parentheses), smallest device heap and agreement with the CuAD adjoint.}
\label{tab:bs}
\resizebox{\textwidth}{!}{%
\begin{tabular}{lccccc}
\toprule
& \multicolumn{3}{c}{Adjoint / primal} & Heap & Agreement \\
Tool & $N=1024$ & $N=4096$ & $N=16384$ & $N=4096$ & with CuAD \\
\midrule
CuAD                            & 3.75 (3.74--3.80) & 3.98 (3.88--7.47) & 3.13 & none & -- \\
Enzyme, \texttt{\_\_restrict\_\_} & 3.25 (3.20--3.26) & 3.59 (3.53--4.21) & 3.68 & none & $4\times10^{-15}$ \\
Enzyme                          & 13.5 (13.4--14.2) & 243 (163--294) & 10.3 & 1~GB & $4\times10^{-15}$ \\
Clad 2.0                        & 62.3 (61.5--66.0) & 4400 (520--8281), 1 of 3 failed & failed & 1~GB & $3\times10^{-15}$ \\
\bottomrule
\end{tabular}}
\end{table}

\subsection{Effect of the accumulation analysis}

Switching off the accumulation analysis isolates its contribution. CuAD then treats \texttt{su} and
\texttt{sv} as ordinary carried scalars and saves both in every iteration, so the stack must be declared with
$2N$ entries. The generated adjoint is correct, and its cost is the same as with the analysis, 3.7 to 4.2 times
the primal, because the saved values are never read and the compiler removes the dead stores together with the
stack. The benefit of the analysis is therefore not speed but the absence of a stack whose size would have to
be chosen at compile time from the problem size. The speed of the CuAD adjoint comes from recomputation under
the no-aliasing convention and from keeping any stack in thread-local memory.

\subsection{Robustness}

Clad terminated with a segmentation fault while differentiating the Euler residual K3. The failure persisted
when the source term was folded into the loop over the three equations or removed altogether, in five of five
repeated builds of each variant. CuAD differentiated the same source without modification, and its tangent
and adjoint pass the tests of Table~\ref{tab:verif}.

% =====================================================================
\section{Why the Jacobian must be exact}
\label{sec:exact}

\subsection{Error amplification}

Let $A=(\partial\R/\partial\bu)^{\mathsf T}$ and let $A+\delta A$ be an approximation of it. If
$\kappa(A)\lVert\delta A\rVert/\lVert A\rVert<1$, the adjoint computed with $A+\delta A$ satisfies
\begin{equation}
\frac{\lVert\delta\bpsi\rVert}{\lVert\bpsi\rVert}\le
\frac{\kappa(A)\,\lVert\delta A\rVert/\lVert A\rVert}{1-\kappa(A)\,\lVert\delta A\rVert/\lVert A\rVert},
\label{eq:bound}
\end{equation}
so a relative matrix error of $10^{-5}$ is harmless only while the condition number stays well below
$10^{5}$. Many GPU solvers compute in single precision and form their Jacobian by finite differences of the
residual kernel. A column computed as $[\R(\bu+\epsilon\mathbf e)-\R(\bu-\epsilon\mathbf e)]/2\epsilon$ has a
truncation error that grows with $\epsilon$ and a cancellation error that grows as $\epsilon$ shrinks, and with
seven significant digits no step makes both small. A Jacobian assembled from the tangent kernel by
Eq.~\eqref{eq:colour} has no truncation error, and its entries carry only the rounding of the tangent
arithmetic.

\subsection{Experiment}

We measure the amplification on the steady Burgers problem of Section~\ref{sec:problems} with $u(0)=1$ and
$u(1)=-0.6$, which has a viscous shock whose position is fixed by the unequal end states. For each case the
discrete steady state is computed by Newton's method in double precision. The tridiagonal Jacobian is then
assembled on the GPU in three ways, each with three colours. The first uses the CuAD tangent of the
single-precision kernel. The second uses single-precision central differences with a step
$\epsilon\max(|u_j|,1)$. The third uses the CuAD tangent of the double-precision kernel and serves as the
reference. For the objective $J=h\sum_iu_i^2$ and a source term $s_i$ added to the residual, the gradient
$\mathrm dJ/\mathrm d\mathbf s$ equals the adjoint vector, and every adjoint system is solved in double
precision. Differences between the gradients therefore come only from the Jacobians. The experiment measures
the error that a single-precision Jacobian alone introduces into a double-precision adjoint; a solver that
also solves its linear systems in single precision would add further errors that are not considered here.

\subsection{Results}

\begin{figure}[ht]
\centering
\includegraphics[width=0.66\textwidth]{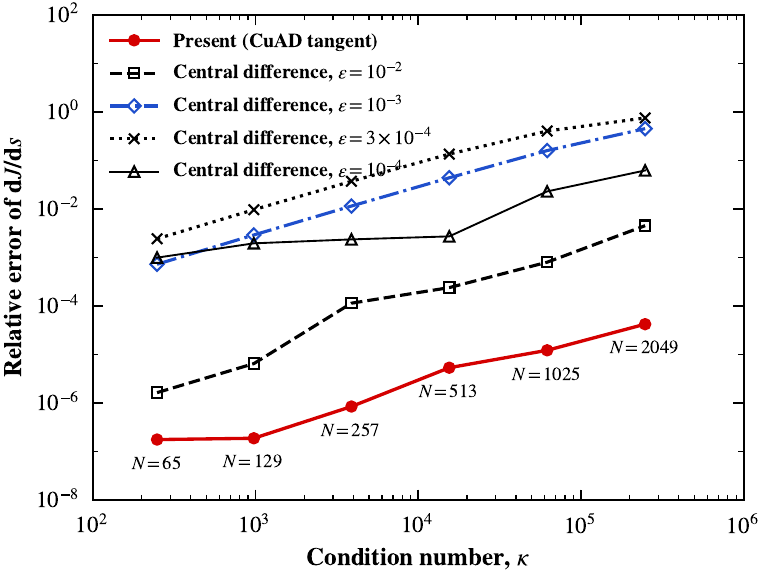}
\caption{Relative error of the adjoint gradient of the steady Burgers problem against the condition number of
its Jacobian, for Jacobians from the CuAD tangent and from central differences, all in single precision. Labels
give the number of grid points.}
\label{fig:burgers}
\end{figure}

Figure~\ref{fig:burgers} shows the relative error of the gradient as the grid is refined from 65 to 2049
points at $\nu=0.02$, which raises the condition number from $2.5\times10^{2}$ to $2.5\times10^{5}$.
Table~\ref{tab:acc} lists the values. The CuAD Jacobian in single precision differs from the reference by
$2.6\times10^{-8}$, which is the rounding of its entries, and its gradient error grows from $1.8\times10^{-7}$
to $4.2\times10^{-5}$. The finite-difference Jacobians differ from the reference by $6\times10^{-7}$ to
$6\times10^{-5}$, one to three orders of magnitude more, and the adjoint solve amplifies this difference. With
the common steps $\epsilon=10^{-3}$ and $3\times10^{-4}$ the gradient error reaches 45 and 75 per cent on the
finest grid, and the cosine between the computed and reference gradients falls to 0.96 and 0.76. The largest
step, $\epsilon=10^{-2}$, is the most accurate of the finite differences, and its error still exceeds that of
the CuAD Jacobian by one to two orders of magnitude. The ordering of the steps follows from the form of the
residual. Equation~\eqref{eq:k1} is quadratic in $\bu$, so a central difference has no truncation error in
exact arithmetic and its whole error is the cancellation of single-precision round-off, which decreases as the
step grows. This property is particular to the Burgers residual, and Section~\ref{sec:exact_euler} repeats the
experiment with a residual that is not polynomial.

\begin{table}[ht]
\centering
\caption{Jacobian and gradient errors for the steady Burgers problem at $\nu=0.02$, all Jacobians in single
precision.}
\label{tab:acc}
\resizebox{\textwidth}{!}{%
\begin{tabular}{rrcccccc}
\toprule
& & \multicolumn{2}{c}{CuAD tangent} & \multicolumn{2}{c}{FD, $\epsilon=10^{-3}$} & \multicolumn{2}{c}{FD, $\epsilon=3\times10^{-4}$} \\
$N$ & $\kappa$ & Jacobian & gradient & Jacobian & gradient & Jacobian & gradient \\
\midrule
65   & $2.5\times10^{2}$ & $2.4\times10^{-8}$ & $1.8\times10^{-7}$ & $1.2\times10^{-5}$ & $7.3\times10^{-4}$ & $4.1\times10^{-5}$ & $2.4\times10^{-3}$ \\
129  & $9.8\times10^{2}$ & $2.7\times10^{-8}$ & $1.9\times10^{-7}$ & $1.1\times10^{-5}$ & $2.9\times10^{-3}$ & $3.7\times10^{-5}$ & $9.7\times10^{-3}$ \\
257  & $3.9\times10^{3}$ & $2.7\times10^{-8}$ & $8.5\times10^{-7}$ & $1.1\times10^{-5}$ & $1.2\times10^{-2}$ & $3.6\times10^{-5}$ & $3.7\times10^{-2}$ \\
513  & $1.6\times10^{4}$ & $2.6\times10^{-8}$ & $5.4\times10^{-6}$ & $1.1\times10^{-5}$ & $4.4\times10^{-2}$ & $3.6\times10^{-5}$ & $1.4\times10^{-1}$ \\
1025 & $6.2\times10^{4}$ & $2.6\times10^{-8}$ & $1.2\times10^{-5}$ & $1.1\times10^{-5}$ & $1.6\times10^{-1}$ & $3.6\times10^{-5}$ & $4.0\times10^{-1}$ \\
2049 & $2.5\times10^{5}$ & $2.6\times10^{-8}$ & $4.2\times10^{-5}$ & $1.1\times10^{-5}$ & $4.5\times10^{-1}$ & $3.6\times10^{-5}$ & $7.5\times10^{-1}$ \\
\bottomrule
\end{tabular}}
\end{table}

In this problem the finite-difference gradients keep the correct sign in every component. The error appears
in the magnitude and in the direction of the gradient, which is what a gradient-based optimiser uses. Whenever
a single-precision solver forms an adjoint from its own Jacobian, an exact Jacobian removes the only error that
the condition number amplifies beyond the rounding of the entries.

\subsection{A transonic airfoil}
\label{sec:exact_euler}

The same comparison was repeated for the two-dimensional Euler residual of Section~\ref{sec:airfoil}, whose
fluxes, pressure switch and boundary conditions are not polynomial in the state. At the converged
double-precision solution of the transonic NACA~0012 on three grids, the Jacobian was assembled from the CuAD
tangent of the single-precision kernels (generated from a single-precision copy of the source), from
single-precision central differences with steps $\epsilon\max(|w_j|,1)$, and from the double-precision tangent
as the reference. Each Jacobian gives a double-precision adjoint and the shape gradients of lift and drag of
Section~\ref{sec:airfoil}. Table~\ref{tab:acc_euler} lists the errors of these gradients.

\begin{table}[ht]
\centering
\caption{Relative errors of the lift and drag shape gradients (eight parameters) of the transonic NACA~0012 when
the Jacobian is formed in single precision; all solves in double precision. $\kappa_1$ is the estimated 1-norm
condition number of the reference Jacobian.}
\label{tab:acc_euler}
\resizebox{\textwidth}{!}{%
\begin{tabular}{lccccc}
\toprule
& & CuAD tangent & FD, $\epsilon=10^{-2}$ & FD, $\epsilon=10^{-3}$ & FD, $\epsilon=10^{-4}$ \\
Grid & $\kappa_1$ & $C_L$ / $C_D$ & $C_L$ / $C_D$ & $C_L$ / $C_D$ & $C_L$ / $C_D$ \\
\midrule
$80\times32$  & $8.9\times10^{6}$ & $8.7\times10^{-7}$ / $3.5\times10^{-7}$ & $1.7\times10^{-2}$ / $9.7\times10^{-3}$ & $1.0\times10^{-3}$ / $1.8\times10^{-3}$ & $2.9\times10^{-3}$ / $1.3\times10^{-3}$ \\
$160\times64$ & $1.9\times10^{7}$ & $3.0\times10^{-6}$ / $1.9\times10^{-6}$ & $1.9\times10^{-2}$ / $1.2\times10^{-2}$ & $3.3\times10^{-3}$ / $1.5\times10^{-3}$ & $1.4\times10^{-3}$ / $1.2\times10^{-3}$ \\
$240\times96$ & $2.0\times10^{7}$ & $1.8\times10^{-6}$ / $4.4\times10^{-7}$ & $4.1\times10^{-2}$ / $2.0\times10^{-2}$ & $5.6\times10^{-4}$ / $1.0\times10^{-3}$ & $2.7\times10^{-3}$ / $6.2\times10^{-4}$ \\
\bottomrule
\end{tabular}}
\end{table}

The single-precision CuAD Jacobian differs from the reference by $6\times10^{-6}$ to $1.4\times10^{-5}$,
mainly because it is evaluated at the state rounded to single precision, and its gradients are accurate to
$4\times10^{-7}$ to $3\times10^{-6}$. The finite-difference Jacobians differ from the reference by
$8\times10^{-4}$ to $1.3\times10^{-2}$, and their gradients carry errors between $6\times10^{-4}$ and
$4\times10^{-2}$, two to four orders of magnitude more. Here the truncation error matters: the largest step is
the least accurate, and no step brings the error below $5\times10^{-4}$. Unlike in the Burgers problem the
gradient error does not grow with the condition number, which is far larger here but leaves the error close to
the relative error of the Jacobian; the bound~\eqref{eq:bound} is not sharp for this system. The direction of
the gradient is kept to a cosine above 0.9996, so the finite-difference gradients still point the right way,
but they carry about three correct digits, while the generated Jacobian in the same precision preserves six.

% =====================================================================
\section{Applications}
\label{sec:apps}

The following applications use the generated kernels inside complete adjoint computations. In each, the
tangent kernel assembles the Jacobian by colouring (Section~\ref{sec:adjoint}), the adjoint kernel supplies
the design sensitivities, and the result is checked against finite differences of the full computation.

\subsection{Adjoint of a viscous shock}
\label{sec:burgersapp}

\begin{figure}[ht]
\centering
\includegraphics[width=\textwidth]{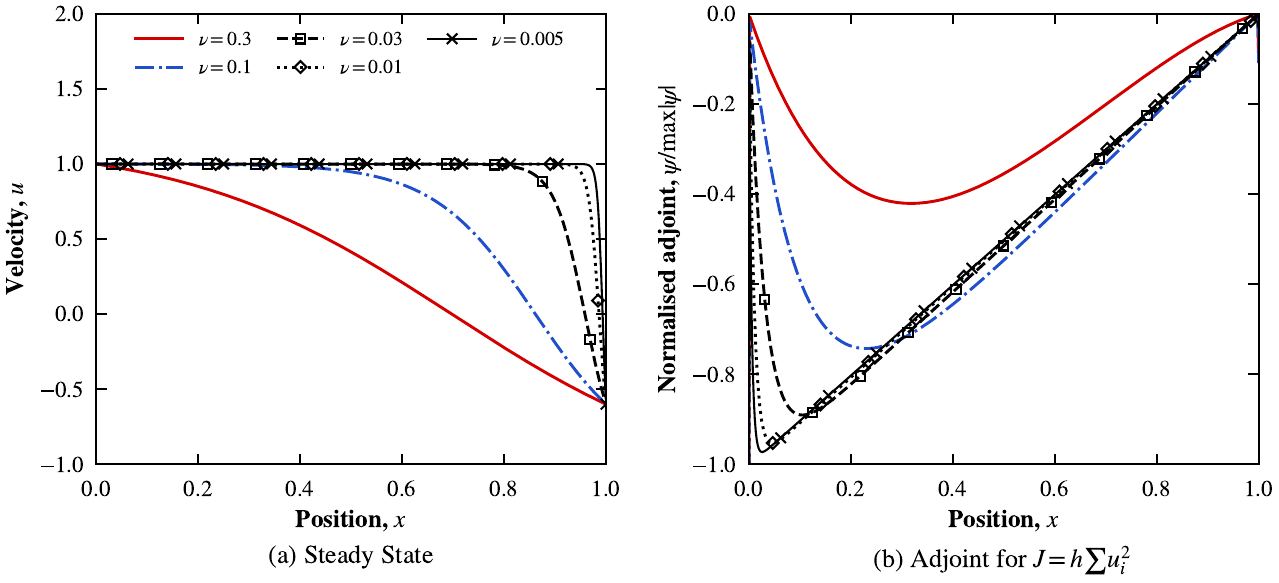}
\caption{Steady viscous Burgers problem with $u(0)=1$ and $u(1)=-0.6$ on 513 points.}
\label{fig:burgersfields}
\end{figure}

Figure~\ref{fig:burgersfields} shows the steady states of the Burgers problem of Section~\ref{sec:exact} for
viscosities from 0.3 to 0.005 on 513 points, together with the adjoint for $J=h\sum_iu_i^2$. As $\nu$
decreases the shock steepens and moves towards the outflow boundary, where it meets the boundary layer formed by
the end state $u=-0.6$. The adjoint, which is also the sensitivity of $J$ to a source term, is largest
upstream of the shock. A source placed in the uniform region upstream changes the state over the whole
domain up to the shock, whereas a source inside the shock is absorbed locally. The adjoint vanishes at both
Dirichlet boundaries and approaches a linear profile in the uniform region as $\nu$ decreases.

\subsection{Sensitivity of a scalar sensor in a cellular vortex}
\label{sec:advdiffapp}

A scalar is released from a Gaussian source of width 0.04 at $(0.25,0.5)$ into the cellular vortex with
stream function $\psi_s=\sin(\pi x)\sin(\pi y)/\pi$ on the unit square, and is measured by the mean
concentration over a disc of radius 0.05 at $(0.75,0.5)$. The diffusivity $\kappa=5\times10^{-3}$ gives a
Péclet number of 200, and the grid of $257^2$ cells gives a cell Péclet number below one, so the central
fluxes of Eq.~\eqref{eq:k2} need no stabilisation. The concentration vanishes on the walls. The Jacobian is
assembled from five tangent evaluations of K2, and the steady state and the adjoint are each one sparse
solve.

\begin{figure}[ht]
\centering
\includegraphics[width=\textwidth]{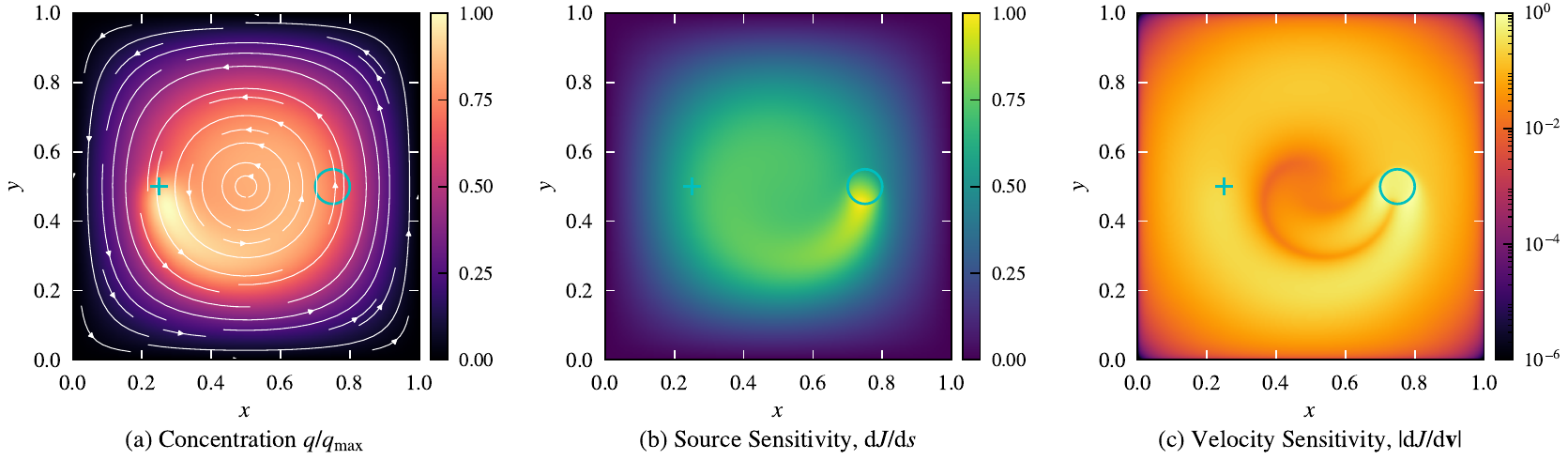}
\caption{Scalar transport in a cellular vortex with source ($+$) and sensor (circle).}
\label{fig:advdiff}
\end{figure}

The adjoint answers two questions at the cost of one solve. The first is where a source would most affect the
sensor: since $\partial\R/\partial\mathbf s=-h^2I$, the sensitivity is $\mathrm dJ/\mathrm d\mathbf s=-h^2\bpsi$,
shown in Figure~\ref{fig:advdiff}(b). It forms a spiral that ends at the sensor, because a source upstream along
a streamline reaches the sensor after part of a revolution while diffusion spreads it across the streamlines.
The second question is how the reading responds to a change of the local velocity. A single call of the
adjoint kernel with the seed $\bpsi$ returns $(\partial\R/\partial a)^{\mathsf T}\bpsi$ and
$(\partial\R/\partial b)^{\mathsf T}\bpsi$, whose magnitude is shown in Figure~\ref{fig:advdiff}(c). It is
largest where the plume passes on its way to the sensor and around the sensor itself. Table~\ref{tab:advdiff}
compares both sensitivities with central differences of the full solution at four cells; they agree to eight
significant digits.

\begin{table}[ht]
\centering
\caption{Sensitivities of the sensor reading from the adjoint and from central differences of the full
solution.}
\label{tab:advdiff}
\begin{tabular}{ccrr}
\toprule
Cell $(i,j)$ & Variable & Adjoint & Finite difference \\
\midrule
(64, 128)  & $s$ & $3.26764715\times10^{-4}$ & $3.26764715\times10^{-4}$ \\
           & $a$ & $2.33358527\times10^{-2}$ & $2.33358526\times10^{-2}$ \\
           & $b$ & $-2.08095376\times10^{-3}$ & $-2.08095377\times10^{-3}$ \\
(128, 64)  & $s$ & $3.71884110\times10^{-4}$ & $3.71884109\times10^{-4}$ \\
           & $a$ & $3.91866105\times10^{-3}$ & $3.91866102\times10^{-3}$ \\
           & $b$ & $2.39821191\times10^{-2}$ & $2.39821190\times10^{-2}$ \\
(192, 160) & $s$ & $2.61940211\times10^{-4}$ & $2.61940212\times10^{-4}$ \\
           & $a$ & $-1.60309926\times10^{-2}$ & $-1.60309926\times10^{-2}$ \\
           & $b$ & $-5.73196990\times10^{-3}$ & $-5.73196987\times10^{-3}$ \\
(100, 200) & $s$ & $2.57161170\times10^{-4}$ & $2.57161169\times10^{-4}$ \\
           & $a$ & $4.22219028\times10^{-3}$ & $4.22219031\times10^{-3}$ \\
           & $b$ & $-1.68409131\times10^{-2}$ & $-1.68409130\times10^{-2}$ \\
\bottomrule
\end{tabular}
\end{table}

\subsection{Inverse design of a transonic nozzle}
\label{sec:nozzleapp}

\begin{figure}[ht]
\centering
\includegraphics[width=\textwidth]{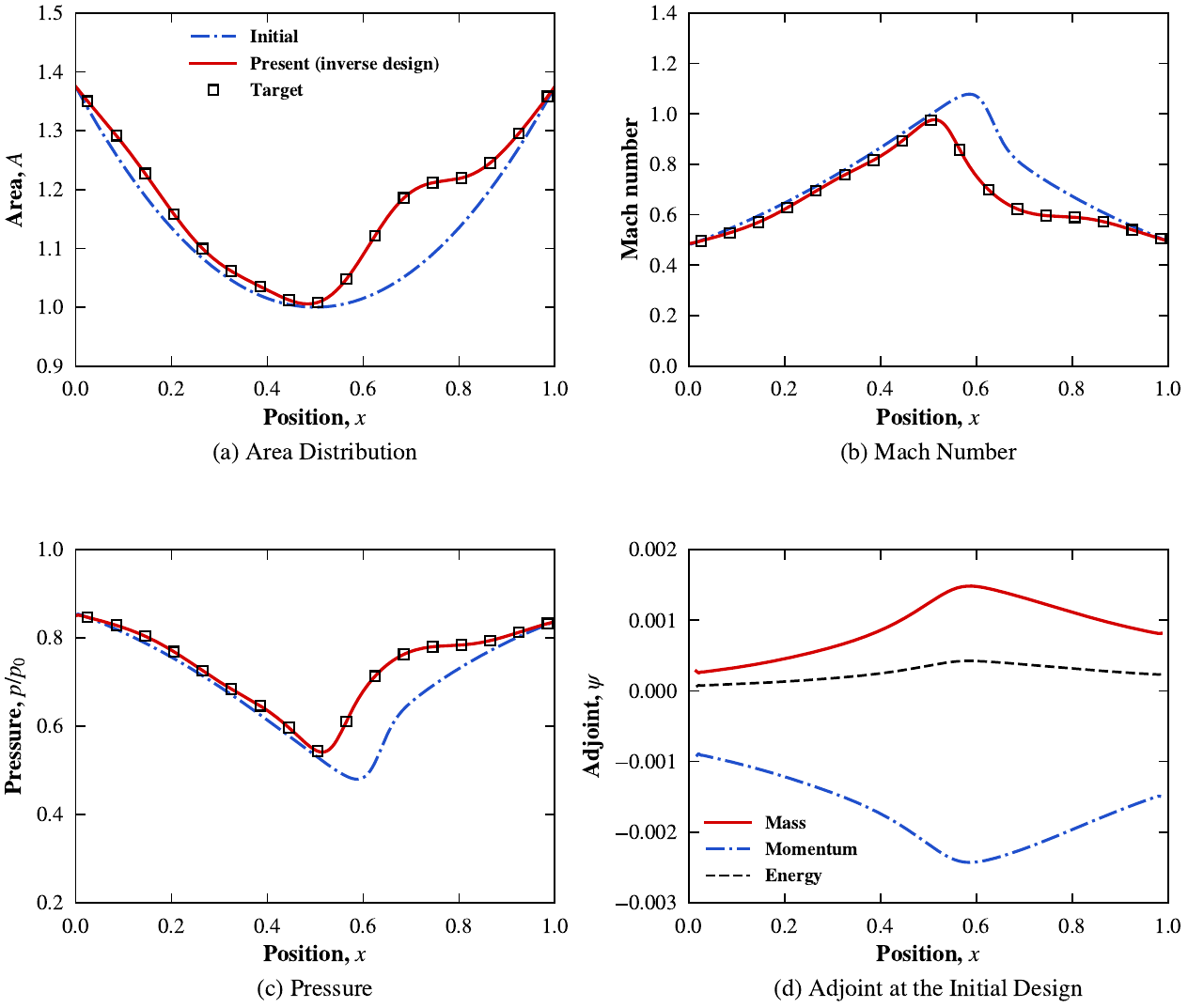}
\caption{Inverse design of a nozzle: area, Mach number and pressure for the initial, target and designed
geometry, and the adjoint at the initial design.}
\label{fig:nozzle}
\end{figure}

\begin{figure}[ht]
\centering
\includegraphics[width=\textwidth]{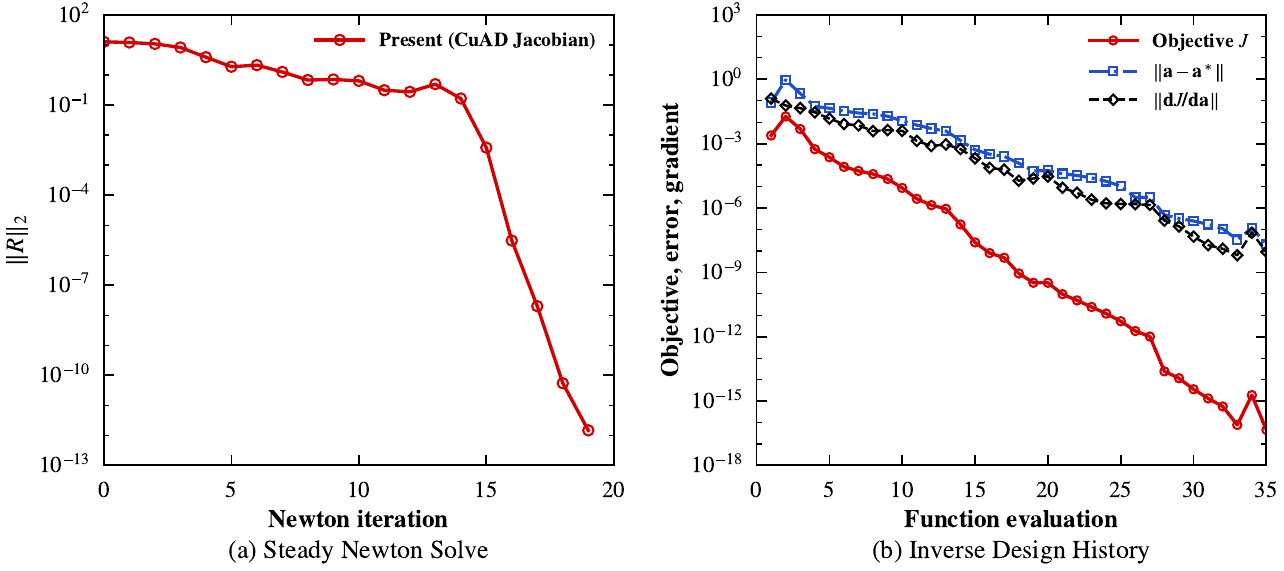}
\caption{Convergence of the steady Newton solve with the Jacobian assembled from the CuAD tangent, and history
of the inverse design.}
\label{fig:nozzledesign}
\end{figure}

The nozzle problem uses the residual K3 on 201 cells with a converging--diverging area
$A_0(x)=1+1.5(x-\tfrac12)^2$. At the inflow the total pressure and total temperature are fixed at $p_0=1$ and
$T_0=1$ and the velocity is extrapolated from the interior. At the outflow the static pressure is fixed at
$p_e=0.84\,p_0$ and the density and velocity are extrapolated. These boundary rows are written in the driver and
differentiated by the complex-step method, while the interior rows and their Jacobian come from the CuAD
kernels. The steady state is found by a pseudo-transient Newton method whose Jacobian is assembled from nine
tangent evaluations of K3, three colours for each of the three conserved variables. Figure~\ref{fig:nozzledesign}
shows that the residual falls from 13 to $1.5\times10^{-12}$ in 20 iterations. The initial flow accelerates
through the throat to a Mach number of 1.08 and recompresses through a smoothed shock near $x=0.6$
(Figure~\ref{fig:nozzle}).

The design problem recovers an unknown area distribution from its pressure. The area is perturbed by eight
sine modes,
\begin{equation}
A(x) = A_0(x) + \sum_{k=1}^{8} a_k\sin(k\pi x),
\end{equation}
which leave the inflow and outflow areas unchanged, and the objective
$J=\tfrac12h\sum_i(p_i-p_i^\ast)^2$ compares the pressure with a target $p^\ast$ computed for a chosen set of
coefficients $\mathbf a^\ast$. The adjoint equation is solved with the transposed Jacobian, and the area
sensitivity $(\partial\R/\partial A)^{\mathsf T}\bpsi$ of the interior rows is one call of the adjoint kernel.
The boundary rows depend on the area of their neighbouring cells, and their contribution is added by the
complex step. Table~\ref{tab:nozzle} compares the gradient with central differences of the full steady solution;
the relative difference of the whole vector is $1.1\times10^{-7}$, the size of the finite-difference error.

\begin{table}[ht]
\centering
\caption{Gradient of the inverse-design objective with respect to the area modes at the initial design.}
\label{tab:nozzle}
\begin{tabular}{crrr}
\toprule
Mode $k$ & Adjoint & Finite difference & Relative difference \\
\midrule
1 & $-8.36192200\times10^{-2}$ & $-8.36192219\times10^{-2}$ & $2.2\times10^{-8}$ \\
2 & $4.67085321\times10^{-2}$  & $4.67085341\times10^{-2}$  & $4.2\times10^{-8}$ \\
3 & $1.68635634\times10^{-2}$  & $1.68635697\times10^{-2}$  & $3.8\times10^{-7}$ \\
4 & $-4.89577739\times10^{-2}$ & $-4.89577768\times10^{-2}$ & $5.8\times10^{-8}$ \\
5 & $3.10116299\times10^{-2}$  & $3.10116233\times10^{-2}$  & $2.1\times10^{-7}$ \\
6 & $-1.23675854\times10^{-3}$ & $-1.23675471\times10^{-3}$ & $3.1\times10^{-6}$ \\
7 & $-4.51467869\times10^{-2}$ & $-4.51467797\times10^{-2}$ & $1.6\times10^{-7}$ \\
8 & $2.80380039\times10^{-2}$  & $2.80379998\times10^{-2}$  & $1.5\times10^{-7}$ \\
\bottomrule
\end{tabular}
\end{table}

With these gradients, the L-BFGS-B method~\cite{byrd1995} drives the objective from $2.4\times10^{-3}$ to
$4.5\times10^{-17}$ in 30 iterations and 35 evaluations, and the area coefficients converge to the target with an
error of $2.1\times10^{-8}$ (Figure~\ref{fig:nozzledesign}). Figure~\ref{fig:nozzle} shows that the design
removes the supersonic pocket: the target area widens the divergent part, so the flow reaches a maximum Mach number
of 0.98 and decelerates without a shock. The adjoint at the initial design is largest in the divergent part where
the initial and target pressures differ most, and its momentum component has the opposite sign of its mass and
energy components.

\subsection{Shape sensitivities of a transonic airfoil}
\label{sec:airfoil}

The last steady application is a complete two-dimensional flow solver. The Euler equations are discretised
by a cell-centred finite-volume method on an O-grid of $160\times64$ cells around the NACA~0012 airfoil, with
the far-field boundary at 20 chords and a first cell height of 0.004 chords. The flux through each face is
the central flux with the scalar artificial dissipation of Jameson, Schmidt and Turkel~\cite{jameson1981}, whose second- and
fourth-difference coefficients are switched by a pressure sensor, $\epsilon_2=k_2\max(\nu_L,\nu_R)$ and
$\epsilon_4=\max(0,k_4-\epsilon_2)$ with $k_2=1/2$ and $k_4=1/64$. Its stencil reaches two cells in each grid
direction. The solver consists of three kernels:
\begin{itemize}
\item \texttt{pad\_state} fills two layers of ghost cells, by reflecting the velocity at the wall and by
characteristic (Riemann-invariant) conditions at the far field, whose branch depends on the sign of the
normal velocity;
\item \texttt{residual} sums the fluxes through the four faces of each cell, with the face normals computed
from the node coordinates;
\item \texttt{wall\_force} gives the pressure force on each wall face, from which lift and drag follow.
\end{itemize}
CuAD differentiates all three kernels, 165 lines of CUDA, without change to the source. Unlike the nozzle of
Section~\ref{sec:nozzleapp}, the boundary conditions are therefore differentiated together with the interior
scheme, and so is the dependence of every term on the mesh. The driver supplies the mesh, the sparse
assembly, the linear solves and the bookkeeping of the shape parameters, in about 90 lines of C and 220 lines
of Python.

\begin{figure}[ht]
\centering
\includegraphics[width=\textwidth]{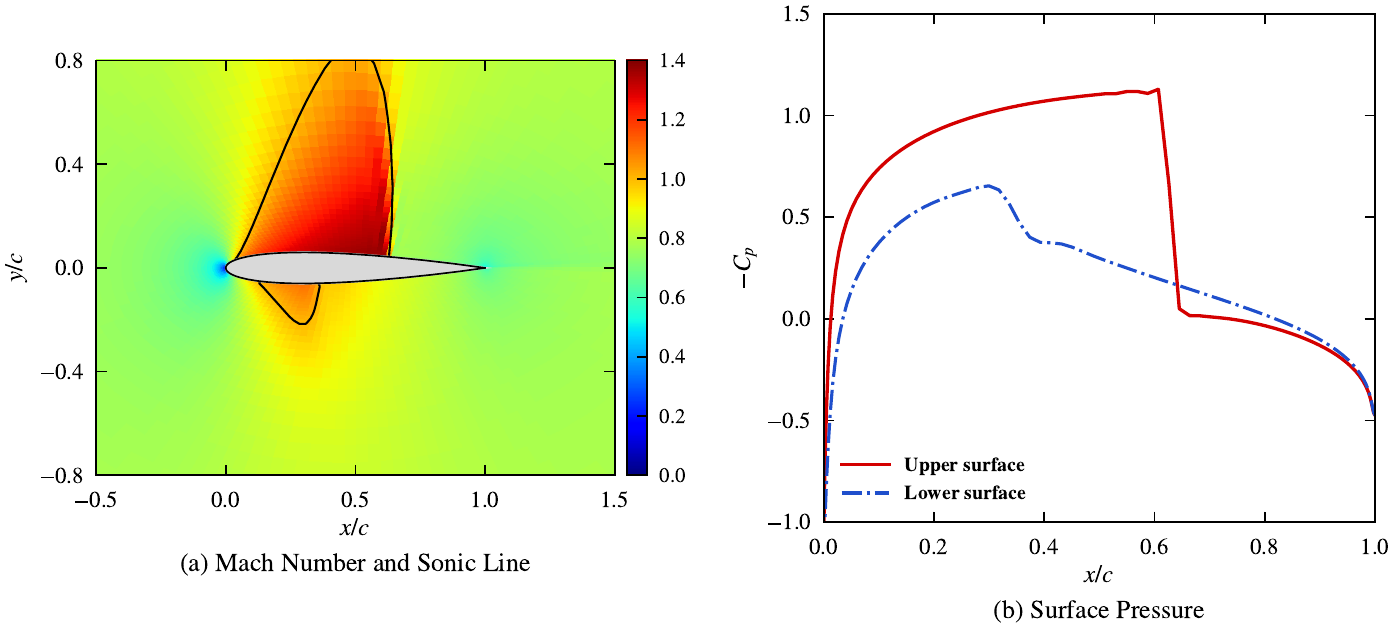}
\caption{Transonic NACA~0012 at $M_\infty=0.8$ and $\alpha=1.25^\circ$ on the $160\times64$ O-grid: Mach number
with the sonic line, and surface pressure.}
\label{fig:airfoil}
\end{figure}

At Mach 0.8 and an angle of attack of $1.25^\circ$, a pseudo-transient Newton method whose Jacobian is
assembled from 100 tangent evaluations (25 colours for the two-cell cross stencil times four variables)
converges in 22 iterations to a residual of $5\times10^{-14}$. The flow (Figure~\ref{fig:airfoil}) has a strong
shock on the upper surface at 62 per cent of the chord and a weak one on the lower surface near 35 per cent,
with $C_L=0.327$ and $C_D=0.0213$. The surface is perturbed by four Hicks--Henne bumps~\cite{hicks1978} on each side, centred at
15, 35, 55 and 75 per cent of the chord, and the mesh follows the surface with a weight that decays to zero at
the far field, so that the nodes depend linearly on the eight parameters.

\begin{figure}[ht]
\centering
\includegraphics[width=\textwidth]{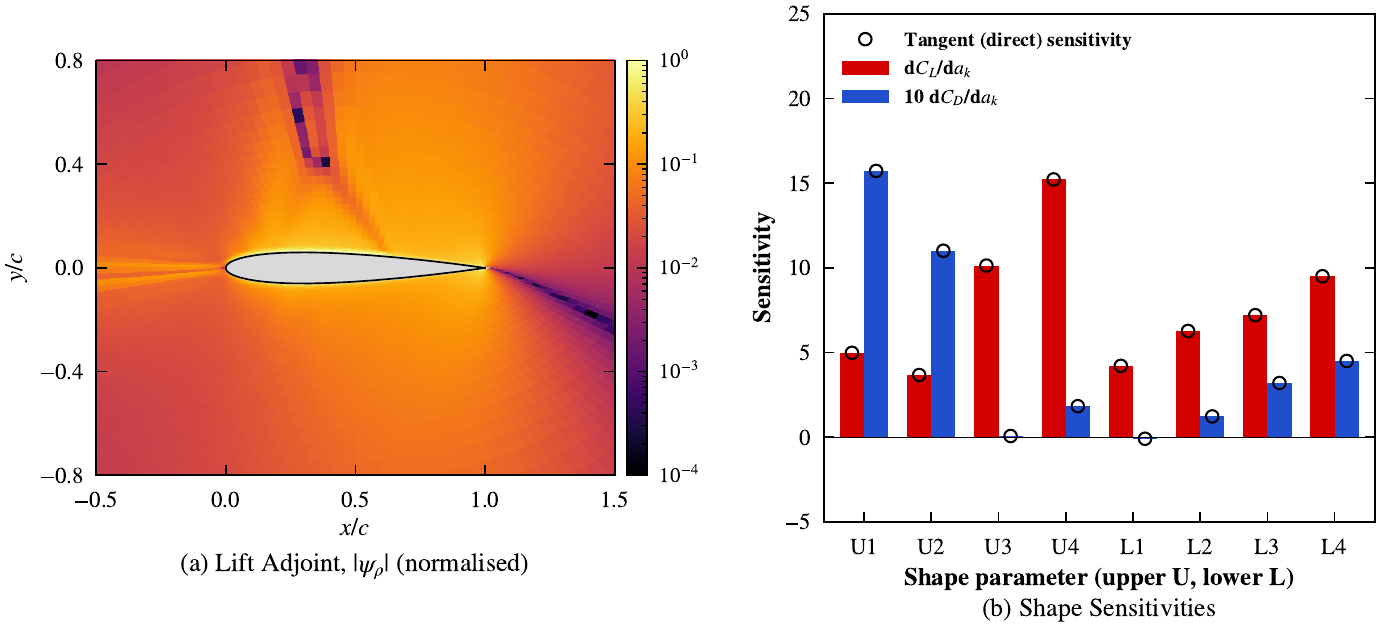}
\caption{Lift adjoint (density component) and the shape sensitivities of lift and drag from the adjoint (bars)
and from the tangent (circles).}
\label{fig:airfoil_adj}
\end{figure}

For each of the lift and drag coefficients one adjoint solve and one call of the adjoint kernels of
\texttt{residual}, \texttt{pad\_state} and \texttt{wall\_force} give the gradient with respect to all node
coordinates, which the linear mesh dependence maps onto the eight parameters (Figure~\ref{fig:airfoil_adj}). The
adjoint identity $\bpsi^{\mathsf T}\partial\R/\partial\mathbf w=-\partial J/\partial\mathbf w$ holds to
$2\times10^{-12}$. As an independent check, the eight sensitivities were also computed in forward mode: a
solve with the Jacobian for the tangent of the residual with respect to each parameter, using the CuAD tangent
kernels. The adjoint and tangent gradients agree to $5\times10^{-14}$ for both coefficients
(Table~\ref{tab:airfoil}). Central differences of the full flow solution approach them as the step is reduced,
with relative differences of $2.3\times10^{-3}$, $1.6\times10^{-4}$ and $3.8\times10^{-5}$ for steps of
$10^{-5}$, $10^{-6}$ and $10^{-7}$. The slow convergence of the differences is expected, because the switches of
the artificial dissipation and the branch of the far-field condition make the residual only piecewise smooth.

The lift is most sensitive to the rear bumps on the upper surface, which move the shock, and the drag to the
bumps near the leading edge of the upper surface, which change the strength of the supersonic region. The lift
adjoint (Figure~\ref{fig:airfoil_adj}a) is largest close to the surface and decays away from it.

On the GPU each assembly of the Jacobian takes 0.14~s, and the adjoint kernels for one gradient take
milliseconds. The steady solution takes 133~s and each adjoint solve 6~s, almost entirely in the sparse
direct factorisation on the host, which an iterative solver on the device would replace in a production
setting.

\begin{table}[ht]
\centering
\caption{Lift and drag sensitivities of the NACA~0012 with respect to the eight shape parameters from the
adjoint, from the tangent (direct) method and from central differences of the full solution with a step of
$10^{-7}$.}
\label{tab:airfoil}
\resizebox{\textwidth}{!}{%
\begin{tabular}{lrrrrr}
\toprule
Parameter & d$C_L$/d$a_k$ (adjoint) & d$C_L$/d$a_k$ (tangent) & d$C_L$/d$a_k$ (FD) & d$C_D$/d$a_k$ (adjoint) & d$C_D$/d$a_k$ (tangent) \\
\midrule
U1 & $4.9891143158$ & $4.9891143158$ & $4.988976$ & $1.5726586312$ & $1.5726586312$ \\
U2 & $3.6768809885$ & $3.6768809885$ & $3.676730$ & $1.1015091750$ & $1.1015091750$ \\
U3 & $10.1385725902$ & $10.1385725902$ & $10.138030$ & $0.0077390686$ & $0.0077390686$ \\
U4 & $15.2251729864$ & $15.2251729864$ & $15.225166$ & $0.1839179863$ & $0.1839179863$ \\
L1 & $4.2195666690$ & $4.2195666690$ & $4.219391$ & $-0.0083624748$ & $-0.0083624748$ \\
L2 & $6.2802783558$ & $6.2802783558$ & $6.280035$ & $0.1234618667$ & $0.1234618667$ \\
L3 & $7.2114677135$ & $7.2114677135$ & $7.211132$ & $0.3206679633$ & $0.3206679633$ \\
L4 & $9.5071288574$ & $9.5071288574$ & $9.506574$ & $0.4508595231$ & $0.4508595231$ \\
\bottomrule
\end{tabular}}
\end{table}

\subsection{Unsteady adjoint of a rolling-up vortex sheet}
\label{sec:vortex}

\begin{figure}[ht]
\centering
\includegraphics[width=\textwidth]{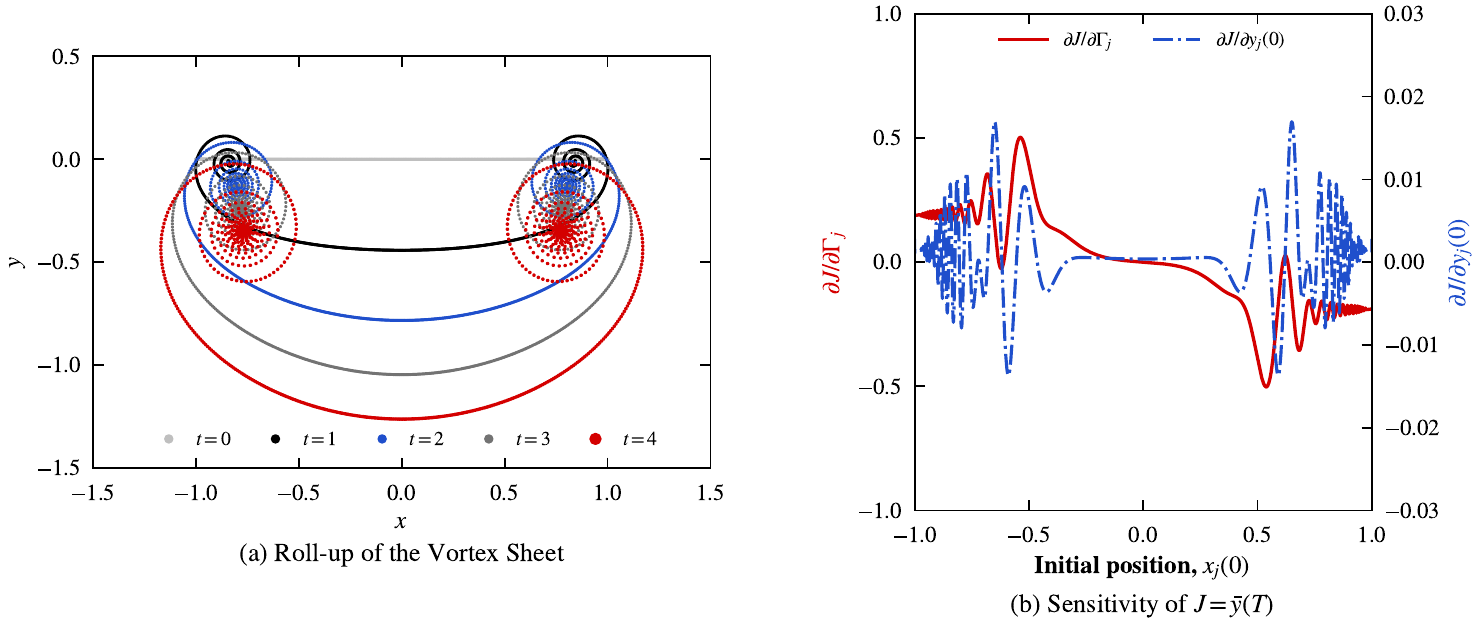}
\caption{Vortex sheet behind an elliptically loaded wing: roll-up and sensitivity of the wake descent.}
\label{fig:vortex}
\end{figure}

The wake behind an elliptically loaded wing rolls up into two tip vortices and descends under its own induced
velocity. Following Krasny~\cite{krasny1987}, we represent the sheet by $N=1024$ point vortices regularised with
$\delta=0.05$~\cite{krasny1986},
placed at $x_j=-\cos(\pi(j-\tfrac12)/N)$ so that they cluster near the tips. Their circulations are the
differences of the elliptic bound circulation $\Gamma(x)=\sqrt{1-x^2}$ between neighbouring panel edges. The
positions are advanced to $T=4$ with Heun's method and $\Delta t=0.01$,
\begin{equation}
\mathbf x^\ast = \mathbf x^n + \Delta t\,\mathbf u(\mathbf x^n),\qquad
\mathbf x^{n+1} = \mathbf x^n + \tfrac12\Delta t\left[\mathbf u(\mathbf x^n)+\mathbf u(\mathbf x^\ast)\right],
\end{equation}
where $\mathbf u$ is the Biot--Savart kernel K4. Figure~\ref{fig:vortex} shows the roll-up of the sheet into
spiral tip vortices and its descent.

The objective is the mean vertical position of the vortices at the final time, $J=\bar y(T)$, a measure of the
wake descent. Its gradient with respect to all 1024 circulations and all initial positions follows from
Eq.~\eqref{eq:unsteady}. In each of the 400 reverse steps, the adjoint of the second stage is one call of the
adjoint kernel at the stored intermediate positions with the seed $\tfrac12\Delta t\,\bar{\mathbf x}^{n+1}$, and the
adjoint of the first stage is a second call at $\mathbf x^n$ whose seed collects both paths through
$\mathbf u(\mathbf x^n)$. The reverse sweep takes about twice as long as the forward integration, since it
evaluates the velocity once and its adjoint twice per step. Table~\ref{tab:vortex} compares the gradient with
central differences of the full time integration; the relative differences lie between $7\times10^{-11}$ and
$2\times10^{-7}$.

\begin{table}[ht]
\centering
\caption{Sensitivity of the wake descent from the unsteady adjoint and from central differences of the full time
integration.}
\label{tab:vortex}
\resizebox{\textwidth}{!}{%
\begin{tabular}{crrrr}
\toprule
$j$ & $\partial J/\partial\Gamma_j$ (adjoint) & $\partial J/\partial\Gamma_j$ (FD) & $\partial J/\partial y_j(0)$ (adjoint) & $\partial J/\partial y_j(0)$ (FD) \\
\midrule
5   & $1.8852071680\times10^{-1}$  & $1.8852071682\times10^{-1}$  & $1.5638553852\times10^{-3}$  & $1.5638554340\times10^{-3}$ \\
100 & $1.9906767355\times10^{-1}$  & $1.9906767340\times10^{-1}$  & $1.9955371952\times10^{-3}$  & $1.9955371799\times10^{-3}$ \\
300 & $3.7638762505\times10^{-2}$  & $3.7638762551\times10^{-2}$  & $-9.9949330243\times10^{-3}$ & $-9.9949329369\times10^{-3}$ \\
512 & $-1.0060143945\times10^{-4}$ & $-1.0060141609\times10^{-4}$ & $3.7744922659\times10^{-4}$  & $3.7744923853\times10^{-4}$ \\
900 & $-1.9177212134\times10^{-1}$ & $-1.9177212152\times10^{-1}$ & $1.7123677499\times10^{-3}$  & $1.7123678586\times10^{-3}$ \\
\bottomrule
\end{tabular}}
\end{table}

The sensitivity to the circulation is antisymmetric because the shed circulations change sign between the two
halves of the sheet. Its largest values appear at about 0.55 of the half-span, the part of the sheet that feeds the
tip vortices during roll-up. Increasing the circulation shed there strengthens the counter-rotating pair and
increases its mutual induction, which accelerates the descent. The sensitivity to the initial height oscillates
near the tips, where the sheet is wound into the spirals and small initial displacements change the winding.

% =====================================================================
\section{Limitations}
\label{sec:limits}

The reverse mode of CuAD reverses counted \texttt{for} loops. Loops written as \texttt{while}, \texttt{do} or
\texttt{switch} statements, and loops that exit with \texttt{break}, are rejected when they touch active data.
Integer locals are not taped, so an integer variable that takes several values is split into one variable per
value. In-place updates of an element of a local array are supported inside loops, while straight-line in-place
updates that an earlier statement reads are not. Conditions are evaluated again in the reverse sweep, which
requires that a branch does not modify its own condition. The rule for atomic updates follows a convention for
thread-local data rather than a proof of the absence of conflicts as in Ref.~\cite{huckelheim2022icpp}. The scalar
parameters of a \texttt{\_\_global\_\_} kernel are passive because they are uniform over the launch, so a quantity
whose derivative is needed is passed as an array.

The absence of a tape in the kernels of this paper rests on two conditions that CuAD assumes rather than
verifies. The first is that a kernel does not overwrite the arrays it reads; a kernel that violates this
convention receives an adjoint that recomputes from overwritten data. The second is the structure of the loops:
a loop that carries scalars other than accumulators from one iteration to the next stores them in every
iteration, and its stack grows with the trip count. The size of the per-thread stack is a compile-time constant
rather than the result of an analysis, and the driver reserves local memory for it in every resident thread. In
the airfoil solver the default stack of 256 entries in the adjoint of \texttt{pad\_state}, which needs only a few,
accounts for most of the 0.8~GB of device memory that the computation occupies. The nozzle of
Section~\ref{sec:nozzleapp} differentiates its boundary rows by the complex-step method in the driver, whereas
the airfoil solver differentiates its boundary conditions as kernels. Finally, all applications solve their
linear systems with a sparse direct method on the host, so their wall-clock times reflect that solver rather
than the generated kernels.

% =====================================================================
\section{Conclusions}
\label{sec:concl}

This work started from the observation that the kernels of finite-difference and finite-volume solvers have a
structure that general GPU differentiation tools do not use. They read their inputs and write their outputs in
separate arrays, and their loops are either short stencils or accumulations. CuAD builds its reverse mode on this
structure. It adopts the no-aliasing convention, recomputes intermediate values instead of storing them,
differentiates accumulations without a tape, keeps any remaining stack in a fixed thread-local array and restricts
atomic updates to memory that several threads share. The adjoints of all kernels studied here need no stored values
and cost two to six times their primal kernels.

The comparison with Clad and Enzyme on one compiler identifies what decides the cost of a GPU adjoint. Enzyme
reaches the cost of CuAD as soon as the source declares the absence of aliasing, so the convention, and not the
particular tool, is what allows the loop to be recomputed. When the convention is not available, values are stored,
and a store in the device heap shared by all threads makes the cost grow by two to three orders of magnitude, vary
from run to run and eventually fail. A source-to-source tool can apply the convention to unmodified kernels and
leaves the result in readable CUDA, which is the practical advantage of CuAD over a compiler-level tool for the
developer of a CUDA flow solver.

The second finding concerns accuracy. A Jacobian assembled from generated tangents is exact up to rounding at the
cost of a fixed number of tangent evaluations. In single precision it kept the adjoint gradients of a stiff Burgers
problem and of a transonic airfoil accurate to $4\times10^{-5}$ and $3\times10^{-6}$, while finite-difference Jacobians
in the same precision gave errors of tens of per cent in the first case and of $5\times10^{-4}$ to $4\times10^{-2}$ in
the second. The size of this error depends on the problem, but its floor is set by the step and cannot be removed.

The applications show that the generated kernels suffice to build complete discrete adjoints. For the transonic
airfoil, the residual, both boundary conditions and the wall force were differentiated without change to the source,
and the shape sensitivities of lift and drag from the adjoint matched those from the tangent to $10^{-13}$. The other
applications recovered a nozzle geometry to eight significant digits and carried an unsteady adjoint through 400 steps
of a vortex-sheet roll-up. Future work will size the per-thread stack by analysis, extend the reverse mode to general
loops, add a proof-based treatment of adjoint writes and replace the host solvers by iterative solvers on the device.

\section*{Code availability}
CuAD is available under the GNU General Public License version~3 at \url{https://github.com/SEUNGHYUN-JOO/CuAD}.
The results in this paper were obtained with release v0.2.1. The kernels, the drivers of all experiments and
applications, the scripts that produce the figures and the raw results are provided with this paper.

% =====================================================================
\appendix
\section{Computational environment}
\label{app:env}

All measurements were made on one workstation with an NVIDIA GeForce RTX~5080 (compute capability 12.0, 16~GB),
under Ubuntu~22.04 in the Windows Subsystem for Linux, with CUDA~12.8. CuAD kernels in
Sections~\ref{sec:verif} and~\ref{sec:apps} were compiled by \texttt{nvcc} with \texttt{-O3 -arch=sm\_120}. For the
comparison of Section~\ref{sec:clad}, every adjoint was compiled by Clang~19.1.7 with \texttt{-O3} for
\texttt{sm\_90} and run through just-in-time compilation of the PTX, with Clad~2.0 and Enzyme~0.0.186 as Clang
plugins. Timings use CUDA events around single launches; Table~\ref{tab:verif} reports the median of 50 launches
and Section~\ref{sec:clad} the median over three processes of the median of five to ten launches. The device heap
for Clad and Enzyme was set with \texttt{cudaDeviceSetLimit} before the first launch. The host linear algebra uses
SciPy's sparse direct solver. The scripts that run every experiment are provided with this paper.
\section{Generated code}
\label{app:listings}

Listing~\ref{lst:k1b} reproduces the complete adjoint of the Burgers kernel K1. The forward sweep repeats the primal
statements, and the reverse sweep consumes the seed \texttt{r\_b[i]}, propagates it through the convective and diffusive
terms and writes the adjoints of the three stencil values with one atomic update each. Listing~\ref{lst:k3d} shows the
beginning of the tangent of the Euler kernel K3. The flux device function is replaced by its tangent companion, and
each spectral-radius call inside a sum is preceded by a call of its companion whose result enters the tangent of the
sum.

\begin{lstlisting}[language=C++,caption={Adjoint of the Burgers kernel K1 as generated by CuAD.},label={lst:k1b}]
__global__ void burgers_res_b(const double * u, double *u_b, const double * f, double *f_b, double * r, double *r_b, double nu, double h, int n)
{
  double nu_b = 0;  double h_b = 0;
  double um;  double um_b = 0;  double uc;  double uc_b = 0;  double up;  double up_b = 0;
  double conv;  double conv_b = 0;  double diff;  double diff_b = 0;
  int i = blockIdx.x * blockDim.x + threadIdx.x;
  if (i > 0 && i < n - 1) {
    um = u[i - 1];  uc = u[i];  up = u[i + 1];
    conv = uc * (up - um) / (2.0 * h);
    diff = nu * (up - 2.0 * uc + um) / (h * h);
    r[i] = conv - diff - f[i];
  }
  if (i > 0 && i < n - 1) {
    double _cuad_b0 = r_b[i];
    r_b[i] = 0;
    conv_b += _cuad_b0;
    diff_b += -(_cuad_b0);
    atomicAdd(&f_b[i], -(_cuad_b0));
    nu_b += ((diff_b)/((h * h)))*((up - 2.0 * uc + um));
    up_b += (nu)*((diff_b)/((h * h)));
    uc_b += (2.0)*(-((nu)*((diff_b)/((h * h)))));
    um_b += (nu)*((diff_b)/((h * h)));
    h_b += (-((diff_b)*(nu * (up - 2.0 * uc + um)))/(((h * h))*((h * h))))*(h);
    h_b += (h)*(-((diff_b)*(nu * (up - 2.0 * uc + um)))/(((h * h))*((h * h))));
    uc_b += ((conv_b)/((2.0 * h)))*((up - um));
    up_b += (uc)*((conv_b)/((2.0 * h)));
    um_b += -((uc)*((conv_b)/((2.0 * h))));
    h_b += (2.0)*(-((conv_b)*(uc * (up - um)))/(((2.0 * h))*((2.0 * h))));
    atomicAdd(&u_b[i + 1], up_b);
    atomicAdd(&u_b[i], uc_b);
    atomicAdd(&u_b[i - 1], um_b);
  }
}
\end{lstlisting}

\begin{lstlisting}[language=C++,caption={Beginning of the tangent of the Euler kernel K3 as generated by CuAD.},label={lst:k3d}]
__global__ void nozzle_res_d(const double * w, const double * w_d, const double * area, const double * area_d, double * res, double * res_d, double eps2, double eps2_d, double h, double h_d, int n)
{
  int i = blockIdx.x * blockDim.x + threadIdx.x;
  if (i > 0 && i < n - 1) {
    double fl_d[3] = {0};  double fc_d[3] = {0};  double fr_d[3] = {0};
    double fl[3], fc[3], fr[3];
    euler_flux_d(w[3 * (i - 1)], w_d[3 * (i - 1)], w[3 * (i - 1) + 1], w_d[3 * (i - 1) + 1],
                 w[3 * (i - 1) + 2], w_d[3 * (i - 1) + 2], area[i - 1], area_d[i - 1], fl, fl_d);
    ...
    double _cuad_c0_d = 0;
    sound_d(w[3 * (i - 1)], w_d[3 * (i - 1)], w[3 * (i - 1) + 1], w_d[3 * (i - 1) + 1],
            w[3 * (i - 1) + 2], w_d[3 * (i - 1) + 2], area[i - 1], area_d[i - 1], _cuad_c0_d);
    double _cuad_c1_d = 0;
    sound_d(w[3 * i], w_d[3 * i], w[3 * i + 1], w_d[3 * i + 1], w[3 * i + 2], w_d[3 * i + 2],
            area[i], area_d[i], _cuad_c1_d);
    double sl_d = (0.5)*((_cuad_c0_d + _cuad_c1_d));
    double sl = 0.5 * (sound(w[3 * (i - 1)], w[3 * (i - 1) + 1], w[3 * (i - 1) + 2], area[i - 1]) +
                       sound(w[3 * i], w[3 * i + 1], w[3 * i + 2], area[i]));
    ...
\end{lstlisting}

\bibliographystyle{unsrtnat}
\bibliography{refs}
\end{document}